\documentclass[11pt]{article}
\usepackage{amsmath}
\usepackage{amsthm}
\usepackage{amssymb}
\usepackage{latexsym}
\usepackage[numbers]{natbib}
\usepackage[dvips]{graphicx,color}
\usepackage{float}
\usepackage{array}
\usepackage{xcolor}
\usepackage{enumerate} 
\usepackage[shortlabels]{enumitem}
\usepackage{titlesec}
\usepackage{multirow}
\usepackage[hyphens]{url}
\usepackage{hyperref}
\usepackage{etoolbox}
\usepackage{tcolorbox}
\usepackage{MnSymbol}
\usepackage{ifthen}

\newtheoremstyle{mystyle}%
{3pt}
{3pt}
{\color{blue}}
{}
{\bfseries\color{blue}}
{.}
{.5em}
{}

\theoremstyle{plain}
\newtheorem{theorem}{Theorem}[section]

\newtheorem{proposition}[theorem]{Proposition}
\newtheorem{corollary}[theorem]{Corollary}

\theoremstyle{definition}

\theoremstyle{remark}

\newtheorem{problem}[theorem]{Problem}

\theoremstyle{mystyle}

\newcommand{\true}{\ensuremath{\mathsf{True}}}
\newcommand{\false}{\ensuremath{\mathsf{False}}}

\newcommand{\eat}[1]{}

\newcommand{\str}[1]{\ensuremath{\mathsf{Str}(#1)}}

\newcommand{\tbf}[1]{\textbf{#1}}
\newcommand{\mc}[1]{\mathcal{#1}}

\makeatletter
\newsavebox\myboxA
\newsavebox\myboxB
\newlength\mylenA

\newcommand*\xoverline[2][0.75]{%
    \sbox{\myboxA}{$\m@th#2$}%
    \setbox\myboxB\null
    \ht\myboxB=\ht\myboxA%
    \dp\myboxB=\dp\myboxA%
    \wd\myboxB=#1\wd\myboxA
    \sbox\myboxB{$\m@th\overline{\copy\myboxB}$}
    \setlength\mylenA{\the\wd\myboxA}
    \addtolength\mylenA{-\the\wd\myboxB}%
    \ifdim\wd\myboxB<\wd\myboxA%
       \rlap{\hskip 0.5\mylenA\usebox\myboxB}{\usebox\myboxA}%
    \else
        \hskip -0.5\mylenA\rlap{\usebox\myboxA}{\hskip 0.5\mylenA\usebox\myboxB}%
    \fi}
    
  \newcommand*\xunderline[2][0.75]{%
    \sbox{\myboxA}{$\m@th#2$}%
    \setbox\myboxB\null
    \ht\myboxB=\ht\myboxA%
    \dp\myboxB=\dp\myboxA%
    \wd\myboxB=#1\wd\myboxA
    \sbox\myboxB{$\m@th\underline{\copy\myboxB}$}
    \setlength\mylenA{\the\wd\myboxA}
    \addtolength\mylenA{-\the\wd\myboxB}%
    \ifdim\wd\myboxB<\wd\myboxA%
       \rlap{\hskip 0.5\mylenA\usebox\myboxB}{\usebox\myboxA}%
    \else
        \hskip -0.5\mylenA\rlap{\usebox\myboxA}{\hskip 0.5\mylenA\usebox\myboxB}%
    \fi}
\makeatother
\newcommand{\ul}[1]{\underline{#1}}

\renewcommand{\str}[1]{\ensuremath{\mathfrak{#1}}}
\newcommand{\anncupdot}{\ensuremath{\ \ul{\cupdot}\ }}

\newcommand{\infcard}{\ensuremath{\mathbf{CN}_{\ge \omega}}}
\newcommand{\card}{\ensuremath{\mathbf{CN}}}
\newcommand{\ord}{\ensuremath{\mathbf{ON}}}
\newcommand{\nat}{\ensuremath{\mathbb{N}}}
\newcommand{\rhohat}{\ensuremath{\hat{\rho}}}
\newcommand{\tower}[1]{\ensuremath{\mathsf{tower}(#1)}}

\newcommand{\ef}{Ehrenfeucht-Fra\"iss\'e}

\newcommand{\kvc}{\ensuremath{k}\textsc{-Vertex cover}}
\newcommand{\kdom}{\ensuremath{k\text{-}\textsc{Dominating Set}}}
\newcommand{\kclique}{\ensuremath{k\text{-}\textsc{Clique}}}

\newcommand{\fv}{Feferman-Vaught}

\newcommand{\tsig}[1]{\ensuremath{\mathrm{T}\Sigma_{#1}}}
\newcommand{\tpi}[1]{\ensuremath{\mathrm{T}\Pi_{#1}}}

\newcommand{\tsiginf}[1]{\ensuremath{\mathrm{T}\Sigma_{#1}}}
\newcommand{\tpiinf}[1]{\ensuremath{\mathrm{T}\Pi_{#1}}}
\newcommand{\tsinf}[1]{\ensuremath{\tsiginf{#1}}}
\newcommand{\tpinf}[1]{\ensuremath{\tpiinf{#1}}}
\newcommand{\linf}[1]{\ensuremath{\mathcal{L}_{#1, \omega}}}

\newcommand{\equivl}{\ensuremath{\equiv_{\mc{L}}}}
\newcommand{\anntau}{\ensuremath{\ul{\tau}}}

\titleformat{\section}[block]{\Large\sc\filcenter}{\thesection.}{5pt}{}
\titleformat{\subsection}[block]{\sc\filcenter}{\thesubsection.}{5pt}{}

\makeatletter
\providecommand{\institute}[1]{
  \apptocmd{\@author}{\end{tabular}
    \par\smallskip
    \begin{tabular}[t]{c}
    #1}{}{}
}
\providecommand{\email}[1]{
  \apptocmd{\@author}{\end{tabular}
    \par\smallskip
    \begin{tabular}[t]{c}
    \texttt{#1}}{}{}
}
\makeatother

\begin{document}

\title{\tbf{Feferman-Vaught Decompositions for\\ Prefix Classes of First Order Logic}\footnote{This research has been supported by the Leverhulme Trust through a Research Project Grant on ``Logical Fractals".}}

\author{\large{Abhisekh Sankaran}\\[3pt]}

\institute{\small{Department of Computer Science and Technology,}\\\small{University of Cambridge, U.K.}}
\email{\small{\{abhisekh.sankaran\}@cl.cam.ac.uk}}
\date{}
\maketitle

\begin{abstract}
The Feferman-Vaught theorem provides a way of evaluating a first order sentence $\varphi$ on a disjoint union of structures by producing a decomposition of $\varphi$ into sentences which can be evaluated on the individual structures and the results of these evaluations combined using a propositional formula. This decomposition can in general be non-elementarily larger than $\varphi$. We show that for first order sentences in prenex normal form with a fixed number of quantifier alternations, such a decomposition, further with the same number of quantifier alternations, can be obtained in time elementary in the size of $\varphi$. We obtain this result as a consequence of a more general decomposition theorem that we prove for a family of infinitary logics we define. We extend these results by considering binary operations other than disjoint union, in particular sum-like operations such as ordered sum and NLC-sum, that are definable using quantifier-free interpretations.
\end{abstract}

\section{Introduction}\label{section:intro}

The Feferman-Vaught theorem~\cite{FV59} is a classic result from model theory that gives a method to evaluate a first order (FO) sentence over a generalized product of structures by reducing it to the evaluation of other first order sentences over the individual structures and the evaluation of a monadic second order (MSO) sentence over an index structure. One of the simplest generalized products is disjoint union and here in case of finitely many structures, one can replace the evaluation of the mentioned MSO sentence, with the evaluation of a propositional formula. One can also stratify the result by the rank of the FO sentence $\varphi$ being evaluated on the disjoint union, that is, one can have the sentences in the alluded ``decomposition" of $\varphi$ to have the same bound on their rank as that for $\varphi$. These results and their generalizations to MSO have a variety of applications in computer science, such as in showing the decidability of theories, satisfiability checking and algorithmic meta-theorems (see~\cite{Mak04} for a survey). 

Computing the Feferman-Vaught decomposition for an FO sentence $\varphi$ over the binary disjoint union of structures (finite or infinite) takes time that is bounded by an $m$-fold exponential in the size of $\varphi$, where $m$ is the rank of $\varphi$~\cite{Mak04}. This runtime is thus non-elementary in the size of $\varphi$, and cannot be improved in general, owing to a non-elementary lower bound for the size of the decomposition over all finite structures (and hence also arbitrary structures)~\cite{DGKS07}.
The time complexity can however be improved by considering special classes of finite structures, such as those of bounded degree, where it takes at most 3-fold exponential time to compute the decomposition if the degree is at least 3, and 2-fold exponential time if the degree is at most 2~\cite{HHS15}. 


In this paper, we take a different approach towards getting faster decompositions, by observing the syntax of the formulae considered. A well-studied normal form for FO sentences is the prenex normal form (PNF).
A prenex sentence is an FO sentence which begins with a string of quantifiers that is followed by a quantifier-free formula. Every FO sentence is equivalent to a prenex sentence and can be brought into such a PNF form in time polynomial in the size of the FO sentence~\cite{Har16}. 
Let $\Sigma_n$ and $\Pi_n$  denote the classes of all PNF sentences that contain $n-1$ alternations of quantifiers (equivalently, $n$ blocks of quantifiers) in the quantifier prefix, and whose leading quantifier is existential and universal respectively. It turns out that various properties of interest in computer science can be expressed using $\Sigma_n$ or $\Pi_n$ sentences for very low values of $n$, indeed with $n$ as just 2. Examples include parameterized problems such as $\kvc, \kclique$ and  $\kdom$ which are all $\Sigma_2$ expressible (more examples can be found in Appendix A of~\cite{San18}). In program verification, the $\Sigma_2$ fragment is called \emph{Effectively Propositional Logic} (EPR) for which there exist practical implementations of DPLL-based decision procedures for checking satisfiability~\cite{EPR1,EPR2,Gul10}. In databases, $\Pi_2$ sentences are the syntactic form of source-to-target dependencies in the data exchange setting, and also of views in data integration~\cite{data-exch,data-integ}. Again, over special classes of structures such as those of bounded degree as aforementioned, every FO sentence is equivalent to a Boolean combination of $\Sigma_2$ sentences. Thus considering a fixed number of quantifier alternations is a  well-motivated restriction. 

Towards the central results of this paper, we consider a ``tree" generalization of $\Sigma_n$ and $\Pi_n$ formulae, that we denote $\tsig{n}$ and $\tpi{n}$. For any FO formula, any root to leaf path in parse tree of the formula can be seen as a word over the quantifier symbols $\exists$ and $\forall$, the logical connectives $\bigwedge, \bigvee$ and $\neg$, the predicate symbols of $\tau$ along with ``=", and a set of variables. We define $\tsig{n}$ as the class of all FO formulae $\psi$ in negation normal form (NNF, where negations appear only at the atomic level), such that the word corresponding to any root to leaf path in the parse tree of $\psi$ has the form $\exists \cdot  (\exists^*\bigwedge \forall^* \bigvee)^*w$ where the number of quantifier alternations in the word is at most $n-1$, and $w$ contains no quantifiers. Likewise for $\tpi{n}$, this word has the form $\forall \cdot  (\forall^*\bigvee \exists^* \bigwedge)^*w$ with at most $n-1$ quantifier alternations and  $w$ as before. Clearly $\tsig{n}$ and $\tpi{n}$ generalize the $\Sigma_n$ and $\Pi_n$ classes of formulae considered in NNF. 


On the semantic front, we consider binary operations on structures, that are defined using quantifier-free interpretations~\cite{Mak04}. Given two structures $\str{A}$ and $\str{B}$, define the \emph{annotated disjoint union} of $\str{A}$ and $\str{B}$ as the disjoint union of these structures in which the elements of the (sub-)universe of $\str{A}$ are labeled with a new unary predicate. We can now define binary operations on inputs $\str{A}$ and $\str{B}$, using quantifier-free scalar interpretations in the annotated disjoint union of $\str{A}$ and $\str{B}$. Here scalar means that the universe defining formula in the interpretation has only one free variable. Such a binary operation is called a  \emph{quantifier-free sum-like} operation. (This is in  contrast with quantifier-free product-like operations like the direct product, that are     definable using quantifier-free non-scalar (or vectorized) interpretations). A number of well-known operations on structures are quantifier-free and sum-like. For example, the disjoint union, the join of two graphs, the ordered sum of structures, the NLC-sum of graphs~\cite{Wan94}, are all quantifier-free sum-like operations. One can consider Feferman-Vaught decompositions of formulae over such operations as a more general setting than over just disjoint union. We can now state one of the three main results of this paper (Theorem~\ref{thm:FO-genop-decomposition}). Below $\tower{n, \cdot}$ denotes the $n$-fold exponential function, and
$\tsig{n}[m]$ and $\tpi{n}[m]$ respectively denote the classes of $\tsig{n}$ and $\tpi{n}$ sentences of quantifier rank at most $m$.


\begin{theorem}\label{thm:main-1-display}
Let $\mc{L}$ be one of the logics $\tsig{n}[m]$ or $\tpi{n}[m]$ where $n, m \ge 0$. Let $\divideontimes$ be a quantifier-free sum-like binary operation on structures whose defining  (quantifier-free) interpretation is $\Xi$. Then for every $\mc{L}$ sentence $\varphi$, there exists a Feferman-Vaught decomposition $D$ for $\varphi$ over $\divideontimes$ consisting of $\mc{L}$ sentences. Further, the decomposition $D$ has size $\tower{n, O((n+1) \cdot |\varphi| \cdot |\Xi|^2)}$, and can be computed in time $\tower{n, O((n+1) \cdot (|\varphi| \cdot |\Xi|^2)^2)}$. 
\end{theorem}
In other words, computing the Feferman-Vaught decomposition of $\varphi$ over $\divideontimes$ has an elementary dependence on the size of $\varphi$ when the number of quantifier alternations in the mentioned  ``tree PNF" form of $\varphi$ is bounded. Further, this decomposition is stratified (in the sense mentioned earlier) by both the rank of $\varphi$ as well as the number of quantifier alternations in the tree PNF form. As a consequence, we obtain that the $\tsig{n}[m]$ theory of the $\divideontimes$-composite of two structures is determined by the $\tsig{n}[m]$ theories of the individual structures. Likewise for the $\tpi{n}[m]$ theory (cf. Corollary~\ref{cor:FO-composition}). Theorem~\ref{thm:main-1-display} is proven by first showing the result for the annotated disjoint union operation (Theorem~\ref{thm:FO-decomposition}), and then transferring the result to general quantifier-free sum-like operations using the defining interpretations of the latter. Using a similar reasoning and as a related result, we show that the number of $\tsig{n}[m]$ or $\tpi{n}[m]$  formulae with a given number of free variables, considered modulo equivalence, is an elementary function of $m$ when $n$ is bounded (cf. Proposition~\ref{prop:size-calculation}). This is in contrast to the non-elementary lower bound for this number for general FO sentences of rank bounded by $m$~\cite[Chapter 3]{Lib13}.

We go further using the arguments involved in showing Theorem~\ref{thm:main-1-display}, to prove a Feferman-Vaught decomposition result for a family of infinitary logics that generalize $\tsig{n}$ and $\tpi{n}$. These logics are obtained by allowing in the (inductive) definition of $\tsig{n}$ and $\tpi{n}$, the number of quantifier alternations to be an arbitrary ordinal, and the arity of the conjunctions and disjunctions to be an arbitrary cardinal. Specifically, for an ordinal $\lambda \ge 0$  and a cardinal $\kappa \ge \omega$, intuitively the logics $\tsinf{\kappa, \lambda}$ and $\tpinf{\kappa, \lambda}$ are the infinitary extensions of $\tsig{n}$ and $\tpi{n}$ respectively, consisting of formulae $\psi$ in NNF  whose parse trees are such that the word corresponding to any leaf to root path is of the form $w \cdot (\bigvee_{< \kappa} \forall^* \bigwedge_{< \kappa} \exists^*)^{< \lambda} \cdot \exists$ if $\psi$ is in $\tsig{\kappa, \lambda}$ and of the form $w \cdot (\bigwedge_{< \kappa} \exists^* \bigvee_{< \kappa} \forall^*)^{< \lambda} \cdot \forall$ if $\psi$ is in $\tpi{\kappa, \lambda}$. Here $\bigwedge_{< \kappa}$ and $\bigvee_{< \kappa}$ denote that the conjunction and disjunction  respectively are each of arity less than $\kappa$, and $(\cdot)^{< \lambda}$ denotes ``less than $\lambda$ many repetitions'' and can be seen an infinitary extension of the usual Kleene star operation (so $(\cdot)^* = (\cdot)^{< \omega}$). Observe that any quantifier block in $\psi$ is only of finite length. We consider formulae in $\tsinf{\kappa, \lambda}$ and $\tpinf{\kappa, \lambda}$ that have only finitely many free variables. It turns out that for $\lambda < \kappa = \omega_1$, essentially the same logics as $\tsinf{\kappa, \lambda}$ and $\tpinf{\kappa, \lambda}$ have already been studied in the literature of the infinitary logic $\mc{L}_{\omega_1, \omega}$. In particular, Ash and Knight define in~\cite[Chapter 6]{AK00}, the logics $\Sigma_\alpha$ and $\Pi_\alpha$ for an ordinal $\alpha$, as the classes of formulae $\varphi(\bar{x})$ that respectively are countable disjunctions of formulae $\exists \bar{u} \psi(\bar{x}, \bar{u})$ where $\psi(\bar{x}, \bar{u})$ belongs to $\Pi_\beta$, and countable conjunctions of formulae $\forall \bar{u} \psi'(\bar{x}, \bar{u})$ where $\psi'(\bar{x}, \bar{u})$ belongs to $\Sigma_\beta$, for $\beta < \alpha$ in each case. It can be seen that $\Sigma_\alpha$ is contained in $\tpinf{\omega_1, \alpha+1}$ and that $\Pi_\alpha$ is contained in $\tsinf{\omega_1, \alpha+1}$. Given that $\Sigma_\alpha$ and $\Pi_\alpha$ (taken over all $\alpha < \omega_1$) are a normal form for formulae of $\mc{L}_{\omega_1, \omega}$, so are $\tsinf{\kappa, \lambda}$ and $\tpinf{\kappa, \lambda}$ (taken over all $\lambda$) when $\kappa = \omega_1$. (We  allow $\lambda$ to be unrestricted since the logics $\tsinf{\kappa, \lambda}$ and $\tpinf{\kappa, \lambda}$ stabilize for $\lambda$ as $\omega_1$ and beyond, as these are both subclasses of $\mc{L}_{\omega_1, \omega}$.) We do not know whether the logics $\tsinf{\kappa, \lambda}$ and $\tpinf{\kappa, \lambda}$ constitute a normal form for $\mc{L}_{\kappa, \omega}$ though this looks plausible.

Towards the decomposition result for the infinitary logics introduced above, extend the $\tower{n, \cdot}$ function to the ``$\lambda$-fold exponential'' function $\tower{\lambda, \cdot}$, and for a cardinal $\mu \ge 0$, let $\tsinf{\kappa, \lambda}[\mu]$ or $\tpinf{\kappa, \lambda}[\mu]$ denote the subclasses of $\tsinf{\kappa, \lambda}$ or $\tpinf{\kappa, \lambda}$ having formulae whose quantifier rank is at most $\mu$. Here the quantifier-rank is defined analogously as for FO, so in particular as the supremum  of the number of quantifiers in any root to leaf path in the parse tree of the infinitary formula. As the second of the main results of this paper (Theorem~\ref{thm:gen-genop-decomposition}), we show the following.

\begin{theorem}\label{thm:main-2-display}
Let $\kappa, \mu$ be cardinals such that $\kappa$ is infinite and $\mu \ge 0$, and let $\lambda \ge 0$ be an ordinal. Define $\rho(\kappa, \lambda)$ as $\tower{\lambda, \kappa}$ if $\kappa > \omega$, and as $\omega$ otherwise. Let $\mc{L}$ be one of the logics $\tsiginf{\kappa, \lambda}[\mu]$ or $\tpiinf{\kappa, \lambda}[\mu]$, and let $\rhohat(\mc{L})$ be the same logic as $\mc{L}$ except that the parameter $\kappa$ is changed to $\rho(\kappa, \lambda)$ (and the parameters $\lambda$ and $\mu$ are left unchanged). Let $\divideontimes$ be a quantifier-free sum-like binary operation on structures. Then for every $\mc{L}$ sentence $\varphi$, there exists a Feferman-Vaught decomposition for $\varphi$ over $\divideontimes$ consisting of $\rhohat(\mc{L})$ sentences. 
\end{theorem}
Thus every $\mc{L}$ sentence $\varphi$ has a Feferman-Vaught decomposition over $\divideontimes$, consisting of sentences with the same structure of quantifier alternations  and the same bound on the rank as $\varphi$. The only difference is that the sentences can become ``width-wise'' larger, in that the arity of the conjunctions and disjunctions can grow but to less than $\rho(\kappa, \lambda)$. As with Theorem~\ref{thm:main-1-display}, Theorem~\ref{thm:main-2-display} is proven by first showing the result for the annotated disjoint union operation (Theorem~\ref{thm:gen-decomposition}) and then transferring the result to quantifier-free sum-like operations using their defining interpretations. As a consequence of Theorem~\ref{thm:main-2-display}, if $\tsinf{\infty, \lambda}[\mu]$ denotes the union of the logics $\tsinf{\kappa, \lambda}[\mu]$ over all $\kappa$, and $\tpinf{\infty, \lambda}[\mu]$ the union of $\tpi{\kappa, \lambda}[\mu]$ over all $\kappa$, we obtain from  Theorem~\ref{thm:main-2-display}  that for $\mc{L}$ that is one of $\tsinf{\infty, \lambda}[\mu]$ or $\tpinf{\infty, \lambda}[\mu]$, and for any given quantifier-free sum-like operation $\divideontimes$, every $\mc{L}$ sentence has a Feferman-Vaught decomposition over $\divideontimes$ consisting of $\mc{L}$ sentences (Theorem~\ref{thm:gen-genop-decomposition}). This further yields as a corollary, that for $\mc{L}$ as mentioned, the $\mc{L}$ theories of two given structures determines the $\mc{L}$ theory of the $\divideontimes$-composite of the structures (Corollary~\ref{cor:gen-composition}). To the best of our knowledge, Feferman-Vaught decompositions and composition results for infinitary logics have not yet been studied in the literature.

Our last main result of the paper goes back to the logics $\tsig{n}$ and $\tpi{n}$, and considers the subclasses of these consisting of formulae in which every quantifier block in the parse tree of the formula contains exactly $k$ quantifiers. These logics are denoted $\tsig{(n, k)}$ and $\tpi{(n, k)}$ respectively. We give an {\ef} (EF) game characterization for equivalence with respect to the defined logics (Theorem~\ref{thm:EF-thm-tree-prefix-game}). This is via an EF game that is a simple variation on the standard  EF game for FO, and that is a ``two-way'' version of the \emph{$(n, k)$-prefix game} defined in~\cite{DS21}. The game arena for the $(n, k)$-prefix game is a \emph{pair} $(\str{A}, \str{B})$ of structures (instead of a set $\{\str{A}, \str{B}\}$ of structures which is the usual case), and the game consists of $n$ rounds. In the odd rounds, the Spoiler picks a $k$-tuple from $\str{A}$ and the Duplicator must respond with a $k$-tuple from $\str{B}$, and in the even rounds, the Spoiler picks a $k$-tuple from $\str{B}$ and the Duplicator must respond with a $k$-tuple from $\str{A}$. The Duplicator wins if the chosen tuples collectively form  a partial isomorphism between $\str{A}$ and $\str{B}$. In~\cite{DS21}, it was claimed  (but not formally shown) that there is a winning strategy for the Duplicator in this game if, and only if, every $\Sigma_{n, k}$ sentence true in $\str{A}$ is also true in $\str{B}$, where a $\Sigma_{n, k}$ sentence is a $\Sigma_n$ sentence in which every quantifier block contains exactly $k$ quantifiers. It turns out that this claim, while it is correct for its "Only if" direction, is incorrect in its "If" direction, as observed in~\cite{FLRV21}. The latter paper gives the example of linear orders $A$ and $B$ of sizes 5 and 4 respectively, that are equivalent with respect to all sentences having at most 3 quantifiers (and hence with respect to $\Sigma_{3, 1}$ sentences), but which are distinguished by the $(3, 1)$-prefix game; specifically the Spoiler has a winning strategy in the $(3, 1)$-prefix game on the pair $(A, B)$. Fortunately, this error in the claimed characterization of~\cite{DS21} does not have any bearing on the main results of~\cite{DS21} since it is the "Only If" direction of the characterization alone that is used for their results. We remedy this situation by providing a correct characterization of the $(n, k)$-prefix game (Theorem~\ref{thm:EF-thm-prefix-game}), by showing that the Duplicator has a winning strategy in this game if, and only if, every $\tsig{n, k}$ sentence true in $\str{A}$ is also true in $\str{B}$. We extend the $(n, k)$-prefix game to its aforementioned two-way version that we call the \emph{$(n, k)$-tree-prefix} game to give a characterization for equivalence with respect to $\tsig{n, k}$ (equivalently with respect to $\tpi{n, k}$). We finally utilize the $(n, k)$-tree-prefix game to show the composition result that the $\tsig{n, k}$ theories of two structures determine the $\tsig{n, k}$ theory of the $\divideontimes$-composite of the structures for any quantifier-free sum-like binary operation $\divideontimes$. Note that this result is incomparable to the composition result mentioned above for $\tsig{n}[m]$ and $\tpi{n}[m]$ since these  classes are  incomparable with $\tsig{(n', k)}$ and $\tpi{(n', k)}$ for all (non-zero) values of $n, n', k$ and $m$. 

\noindent \vspace{2pt}\textbf{Paper Organization:} In Section~\ref{section:prelims}, we introduce terminology and notation, and formally define the classes $\tsinf{\kappa, \lambda}$ and $\tpinf{\kappa, \lambda}$ and their finitary counterparts $\tsig{n}$ and $\tpi{n}$. In Section~\ref{section:gen-decomposition}, we prove the decomposition result for $\tsinf{\kappa, \lambda}$ and $\tpinf{\kappa, \lambda}$ formulae over the annotated disjoint union operation (Theorem~\ref{thm:gen-decomposition}) and present its implications for the mentioned classes and also $\tsinf{\infty, \lambda}$ and $\tpinf{\infty, \lambda}$ for quantifier-free sum-like operations. In Section~\ref{section:FO-decomposition}, we utilize Theorem~\ref{thm:gen-decomposition} to prove the decomposition result for $\tsig{n}$ and $\tpi{n}$ formulae over the annotated disjoint union and other quantifier-free sum-like operations, along with showing that the decompositions have sizes, and can be obtained in time, bounded by an elementary function of the sizes of the input sentences when $n$ is fixed. In Section~\ref{section:EF-game}, we give a characterization of the $(n, k)$-prefix game defined in~\cite{DS21}, and use it to give an EF game characterization for equivalence with respect to $\tsig{(n, k)}$. We finally conclude in Section~\ref{section:conclusion} presenting various directions for future work.

\noindent \textbf{Related work:} It is known that bounding the number of quantifier alternations allows obtaining finite automata for MSO sentences over words, in elementary time~\cite{Tho97}, in contrast with general non-elementary lower bounds in this context~\cite{MS73}. The same restriction on Presburger arithmetic again yields faster decision procedures~\cite{RL78,Haa14}. Finally, the two variable fragment of FO also  admits an elementary (doubly exponential) Feferman-Vaught decomposition for disjoint union~\cite{GJL15}.

\section{Notation and terminology}\label{section:prelims}
We assume the reader is familiar with the standard syntax and semantics of FO~\cite{Lib13}. Let  $\mathbb{N}$ be the set of all natural numbers (including 0), and $\omega$ denote its cardinality. We also use $\omega$ to denote the first infinite ordinal. For $n \in \mathbb{N}$, we let $[n]$ denote the set $\{1, \ldots, n\}$. We will be concerned in this paper with only finite relational vocabularies $\tau$, that is finite vocabularies $\tau$ containing only relation symbols. For a cardinal $\kappa \ge \omega$, the logic $\linf{\kappa}$ is the extension of FO obtained by allowing conjunctions and disjunctions to have  arity $< \kappa$ instead of these connectives being just binary. Specifically, an $\linf{\kappa}$ formula over $\tau$ in NNF is a formula with finitely many free variables that is built up from atomic formulae of the form $R(x_1, \ldots, x_k)$ and $x_1 = x_2$ and the negations of these, where $R$ is a $k$-ary predicate symbol in $\tau$ and $x_1, \ldots, x_k$ are variables, using existential and universal quantification over finitely many (equivalently single) variables, and conjunctions and disjunctions of arity $< \kappa$. It is easy to see that $\linf{\omega}$ is exactly FO. The logic $\linf{\infty}$ is defined as $\linf{\infty} = \bigcup_{\kappa \ge \omega} \linf{\kappa}$.  

\vspace{2pt} \noindent \tbf{1. The logics \tsinf{\kappa, \lambda} and \tpinf{\kappa, \lambda}}: For a cardinal $\kappa \ge \omega$ and an ordinal $\lambda \ge 0$, we define the subclasses $\tsiginf{\kappa, \lambda}$ and $\tpiinf{\kappa, \lambda}$ of $\linf{\kappa}$ over a vocabulary $\tau$, via simultaneous induction over $\lambda$ as follows.
\begin{itemize}
    \item For the base case of $\lambda = 0$, the classes $\tsiginf{\kappa, \lambda}$ and $\tpiinf{\kappa, \lambda}$ are both equal to the class of all quantifier-free FO formulae over $\tau$ in NNF. So this class is built up from atomic formulae of the form $R(x_1, \ldots, x_k)$ and $x_1 = x_2$ and the negations of these, where $R$ is a $k$-ary predicate symbol in $\tau$ and $x_1, \ldots, x_k$ are variables, using binary conjunctions and disjunctions. 
    \item Inductively assume $\tsiginf{\kappa, \lambda'}$ and $\tpiinf{\kappa, \lambda'}$ have been defined for all $\lambda' < \lambda$. Then:
\begin{itemize}
    \item A $\tsinf{\kappa, \lambda}$ formula is a $\tsinf{\kappa, \lambda, r}$ formula for some $r \in \nat$ where: (i) a formula is in $\tsinf{\kappa, \lambda, 0}$ if it has finitely many free variables, and is of the form $\bigwedge_{i \in I} \gamma_i$ where $I$ is an index set of cardinality $< \kappa$ and $\gamma_i$ is a $\tpiinf{\kappa, \lambda''}$ formula for $\lambda'' < \lambda$; (ii) a formula is in $\tsinf{\kappa, \lambda, r}$ for $r > 0$ if it either is $\tsinf{\kappa, \lambda, r-1}$ formula, or is of the form $\exists y \varphi_1$ where $\varphi_1$ is a $\tsinf{\kappa, \lambda, r-1}$ formula.
    \item A $\tpinf{\kappa, \lambda}$ formula is a $\tpinf{\kappa, \lambda, r}$ formula for some $r \in \nat$ where: (i) a formula is in $\tpinf{\kappa, \lambda, 0}$ if it has finitely many free variables, and is of the form $\bigvee_{i \in I} \gamma_i$ where $I$ is an index set of cardinality $< \kappa$ and $\gamma_i$ is a $\tsinf{\kappa, \lambda''}$ formula for $\lambda'' < \lambda$; (ii) a formula is in $\tpinf{\kappa, \lambda, r}$ for $r > 0$ if it either is a $\tpinf{\kappa, \lambda, r-1}$ formula, or is of the form $\forall y \varphi_1$ where $\varphi_1$ is a $\tpinf{\kappa, \lambda, r-1}$ formula.
\end{itemize}
\end{itemize}

Define the classes $\tsinf{\kappa, \infty}, \tpinf{\kappa, \infty}, \tsiginf{\infty, \lambda},  \tpiinf{\infty, \lambda}, \tsiginf{\infty, \infty}$ and $\tpiinf{\infty, \infty}$ as: 
\begin{equation*}
    \begin{split}
        \tsiginf{\kappa, \infty} & = \bigcup_{\lambda \ge 0} \tsinf{\kappa, \lambda}\\
        \tsiginf{\infty, \lambda} & = \bigcup_{\kappa \ge \omega} \tsinf{\kappa, \lambda}\\
        \tsiginf{\infty, \infty} & = \bigcup_{\lambda \ge 0} \tsinf{\infty, \lambda}\\
    \end{split}
    \quad \quad
    \begin{split}
        \tpinf{\kappa, \infty} & = \bigcup_{\lambda \ge 0} \tpinf{\kappa, \lambda}\\
        \tpinf{\infty, \lambda} & = \bigcup_{\kappa \ge \omega} \tpinf{\kappa, \lambda}\\
        \tpinf{\infty, \infty} & = \bigcup_{\lambda \ge 0} \tpinf{\infty, \lambda}\\
    \end{split}
\end{equation*}

We make various observations about the classes defined above. Firstly, one can see using a simple induction that for $\lambda' < \lambda$, the classes $\tsinf{\kappa, \lambda'}$ and $\tpinf{\kappa, \lambda'}$ are both contained inside each of $\tsinf{\kappa, \lambda}$ and $\tpinf{\kappa, \lambda}$. The classes $\tsinf{\kappa, \lambda}$ and $\tpinf{\kappa, \lambda}$ are incomparable. The negation of $\tsinf{\kappa, \lambda}$ formula is equivalent to a $\tpinf{\kappa, \lambda}$ formula and the negation of $\tpinf{\kappa, \lambda}$ formula is equivalent to a $\tsinf{\kappa, \lambda}$ formula. Also every formula in $\tsiginf{\kappa, \lambda}$  or $\tpiinf{\kappa, \lambda}$ has only finitely many free variables (and hence so does any of its sub-formulae). All these facts are true with $\kappa$ or $\lambda$ or both substituted with $\infty$.

The classes $\tsiginf{\omega, \lambda}$ and $\tpinf{\omega, \lambda}$ are classes of FO formulae and can be seen to constitute a normal form for FO. (In fact, they constitute a normal form already with $\lambda$ finite.) The same is the case for the classes $\tsiginf{\omega_1, \lambda}$ and $\tpinf{\omega_1, \lambda}$ with respect to the logic $\linf{\omega_1}$. This is seen by noting that the mentioned classes subsume the classes $\Sigma_\alpha$ and $\Pi_\alpha$ for an ordinal $\alpha < \omega_1$ that are considered in the context of $\linf{\omega_1}$ and constitute a normal form for this logic (see~\cite[Chapter 6]{AK00}). Indeed 
$\Sigma_\alpha = \tpinf{\omega_1, \alpha+1, 0} \subseteq \tpinf{\omega_1, \alpha+1}$ and $\Pi_\alpha = \tsinf{\omega_1, \alpha+1, 0} \subseteq \tsinf{\omega_1, \alpha+1}$.
We do not know whether $\tsinf{\kappa, \lambda}$ and $\tpinf{\kappa, \lambda}$ constitute a normal form for $\linf{\kappa}$ but this question does not concern us in this paper. (We discuss this question though in Section~\ref{section:conclusion}).


The rank of a formula $\varphi \in \{\tsiginf{\kappa, \lambda}, \tpiinf{\kappa, \lambda}\}$ can be defined analogously as in the case of FO. Specifically, the rank of $\varphi$ is the supremum of the  number of quantifiers appearing in any root-to-leaf path in the parse tree of $\varphi$. For $\mc{L} \in \{\tsinf{\kappa, \lambda, n}, \tpinf{\kappa, \lambda, n}, \tsinf{\kappa, \lambda}, \tpinf{\kappa, \lambda}\}$ (and with one or both of $\kappa, \lambda$ also taking on the value $\infty$), let $\mc{L}[\mu]$ denote the classes of all $\mc{L}$ formulae of rank $\leq \mu$ for a cardinal $\mu \ge 0$. Observe that the negation of a $\tsinf{\kappa, \lambda}[\mu]$ sentence is a equivalent to a $\tpinf{\kappa, \lambda}[\mu]$ sentence and the negation of a $\tpinf{\kappa, \lambda}[\mu]$ sentence is a equivalent to a $\tsinf{\kappa, \lambda}[\mu]$ sentence; likewise with $\tsinf{\kappa, \lambda, n}[\mu]$ and $\tpinf{\kappa, \lambda}[\mu]$ in place of $\tsinf{\kappa, \lambda}[\mu]$ and $\tpinf{\kappa, \lambda}[\mu]$. Also $\tsiginf{\kappa, \lambda}[\mu] = \bigcup_{r \in \mathbb{N}} \tsiginf{\kappa, \lambda, r}[\mu]$ and $\tpinf{\kappa, \lambda}[\mu] = \bigcup_{r \in \mathbb{N}} \tpinf{\kappa, \lambda, r}[\mu]$.


\tbf{The logics $\tsig{n}$ and $\tpi{n}$}: Considering applications in computer science as mentioned in Section~\ref{section:intro}, of particular interest to us are the classes $\tsinf{\kappa, \lambda}$ and $\tpinf{\kappa, \lambda}$ when $\lambda < \kappa = \omega$. As mentioned earlier, these classes already constitute a normal form for FO. We denote these classes respectively as $\tsig{n}$ and $\tpi{n}$ with $n$ playing the role of $\lambda$. The subclasses of these consisting of sentences of rank at most $m$ will be denoted $\tsig{n}[m]$ and $\tpi{n}[m]$ respectively.






\vspace{2pt} \noindent \tbf{2. Interpretations}: We recall the model-theoretic notion of interpretations from the literature~\cite{Hod93}, in particular its special case where the formulae are quantifier-free and contain no parameters. For vocabularies $\tau, \sigma$, a \emph{quantifier-free scalar $(\tau, \sigma)$-interpretation} $\Xi$, or simply a $(\tau, \sigma)$-interpretation $\Xi$, is a tuple  $(\xi_U(x),$ $ (\xi_R(\bar{y}_R))_{R \in \sigma})$ of quantifier-free FO formulas in NNF such that $|\bar{y}_R| = \text{ar}(R)$ where $|\bar{y}_R|$ denotes the lengths of $\bar{y}_R$, and $\text{ar}(R)$ denotes the arity of $R$. Given a $\tau$-structure $\str{A}$, the $(\tau, \sigma)$-interpretation $\Xi$ can be seen to define a $\sigma$-structure $\str{B} = \Xi(\str{A})$ as follows: (i) The universe of $\str{B}$ is given by $B = \xi_U(\str{A}) = \{ a \mid a~\mbox{is an element of}~\str{A}~\mbox{such that}~\str{A} \models \xi_U(a)\}$; (ii) A relation $R \in \sigma$ is interpreted in $\str{B}$ as $R^{\str{B}} = \xi_R(\str{A}) \cap B^{\text{ar}(R)}$ where $\xi_R(\str{A}) = \{ \bar{a} \mid \bar{a}~\mbox{is a}~|\bar{y}_R|\mbox{-tuple from}~\str{A}~\mbox{such that}~\str{A} \models \xi_R(\bar{a})\}$. We say $\str{B}$ is \emph{$\Xi$-interpreted} in $\str{A}$, or simply, interpreted in $\str{A}$. Thus $\Xi$ defines a function from any given class of $\tau$-structures, to $\sigma$-structures. The function is isomorphism-preserving, that is isomorphic $\tau$-structures are mapped to isomorphic $\sigma$-structures. Where it is clear from context, we refer to the function also as a $(\tau, \sigma)$-interpretation. If $\tau$ and $\sigma$ are clear from context, then we call both $\Xi$  and the function it defines, as simply an interpretation. As an example, if $\tau = \sigma = \{E\}$ where $E$ is a binary relation symbol, and $\Xi = (\xi_U(x), \xi_E(x, y))$ where $\xi_U(x) := \true$ and $\xi_E(x, y) = \neg E(x, y)$, then  the function defined by $\Xi$ on undirected graphs is exactly graph complementation. 

One can utilize the mechanism of interpretations as defined above to not just construct unary operations on structures as seen above, but also binary operations. To be able to do so, we first define the \emph{annotated} disjoint union of given $\tau$-structures $\str{A}_1$ and $\str{A}_2$. Let $\str{A}_2'$ be an isomorphic copy of $\str{A}_2$ with universe disjoint from that of $\str{A}_1$. Recall that the disjoint union of $\str{A}_1$ and $\str{A}_2$, denoted $\str{A}_1 \cupdot \str{A}_2$, is the $\tau$-structure defined up to isomorphism as the structure whose universe is the union of the universes of $\str{A}_1$ and $\str{A}_2'$, and in which every predicate of $\tau$ is interpreted as the union of its interpretations in $\str{A}_1$ and $\str{A}_2'$. The \emph{annotated} disjoint union of $\str{A}_1$ and $\str{A}_2$, denoted $\str{A}_1 \anncupdot \str{A}_2$, is the structure defined up to isomorphism by expanding $\str{A}_1 \cup \str{A}_2'$ with a new unary predicate not in $\tau$, that is interpreted as the universe of $\str{A}_1$. Formally, $\str{A}_1 \anncupdot \str{A}_2$ is (up to isomorphism) a $\anntau$-structure $\str{B}$ for $\anntau = \tau \cupdot \{P\}$ and $P$ a unary predicate (not in $\tau$), such that the $\tau$-reduct of $\str{B}$ is $\str{A}_1 \cupdot \str{A}_2'$, and the interpretation of $P$ in $\str{B}$ is the universe of $\str{A}_1$. We can now utilize the annotated disjoint union to define binary operations on $\tau$-structures. In particular, each $(\anntau, \tau)$-interpretation $\Xi$ defines a binary operation $\divideontimes$ on $\tau$-structures given by $\str{A}_1 \divideontimes \str{A}_2 = \Xi(\str{A}_1 \anncupdot \str{A}_2)$. We call $\divideontimes$ a \emph{quantifier-free sum-like binary  operation} on $\tau$-structures, and call $\Xi$ as its \emph{quantifier-free definition}. Following are some well-known binary operations on structures that are quantifier-free and sum-like. 
\begin{enumerate}
    \item Disjoint union: A quantifier-free definition for this operation is $\Xi = (\xi_U(x),$ $ (\xi_R(\bar{y}_R))_{R \in \tau})$ where $\xi_U(x) := \true$ and $\xi_R(\bar{y}_R) := R(\bar{y}_R)$.
    
    
    \item Ordered sum: Here $\tau$ is the vocabulary of ordered structures, and so is of the form $\tau = \{\leq\} \cupdot \tau_1$ where $\leq$ is a binary predicate that is interpreted as a total linear order in an ordered $\tau$-structure. The ordered sum of ordered $\tau$-structures $\str{A}_1$ and $\str{A}_2$ is constructed by taking the disjoint union of the structures, and extending the interpretation of $\leq$ to a total order, by adding the pairs $(a_1, a_2)$ such that $a_1 \in \str{A}_1$ and $a_2 \in \str{A}_2$. 
    Then a quantifier-free definition of the ordered sum is $\Xi = (\xi_U(x), (\xi_R(\bar{y}_R))_{R \in \tau})$ where (i) $\xi_U(x) := \true$; (ii) $\xi_R(\bar{y}_R) := R(\bar{y}_R)$ for $R \in \tau_1$; and (iii)
    $\xi_\leq(y_1, y_2) := (y_1 \leq y_2) \vee (P(y_1) \wedge \neg P(y_2))$.
    \item NLC-sum: Here $\tau$ is the vocabulary of labeled undirected graphs, so is of the form $\tau = \{E \} \cupdot \tau_1$ where $\tau_1 = \{Q_1, \ldots, Q_r\}$ and $Q_i$ is a unary predicate for $i \in [r]$. An $[r]$-labeled graph is a $\tau$-structure whose $\{E\}$-reduct is an undirected graph, and in which the interpretations of the $Q_i$s form a partition of the vertex set of the graph (allowing empty parts). The NLC-sum operation is specified using a set $S \subseteq [r]^2$. It takes as input two $[r]$-labeled graphs  $G_1$ and $G_2$, creates their disjoint union, and adds edges between vertices $u \in G_1$ and $v \in G_2$ such that $G_1 \models Q_i(u)$ and $G_2 \models Q_j(v)$ where $(i, j) \in S$. This operation can then be seen to have a quantifier-free definition  $\Xi = (\xi_U(x), (\xi_R(\bar{y}_R))_{R \in \tau})$ where (i) $\xi_U(x) := \true$; (ii) $\xi_{Q_i}(y) := Q_i(y)$ for $i \in [r]$; and (iii) $\xi_E(y_1, y_2) := E(y_1, y_2) \vee \bigvee_{(i, j) \in S} (P(y_1) \wedge Q_i(y_1) \wedge \neg P(y_2) \wedge Q_j(y_2))$.
\end{enumerate}

Given a $(\tau, \sigma)$-interpretation $\Xi = (\xi_U(x), (\xi_R(\bar{y}_R))_{R \in \sigma})$ and an $\mc{L}$ formula $\varphi(\bar{z})$  over $\sigma$ for $\mc{L} \in \{\tsinf{\kappa, \lambda}, \tpinf{\kappa, \lambda}\}$, let $\Xi(\varphi)$ denote the $\mc{L}$ formula over $\tau$ defined inductively over the structure of $\varphi(\bar{z})$ as follows.
\begin{enumerate}
    \item If $\varphi(\bar{z}) := R(z_1, \ldots, z_r)$ for $R \in \sigma \cup \{=\}$, then  $\Xi(\varphi) := \xi_R(z_1, \ldots, z_r) \wedge \bigwedge_{i \in [r]} \xi_U(z_i)$.
    \item If $\varphi(\bar{z}) := \neg R(z_1, \ldots, z_r)$ for $R \in \sigma \cup \{=\}$, then $\Xi(\varphi) := \neg \xi_R(z_1, \ldots, z_r) \wedge \bigwedge_{i \in [r]} \xi_U(z_i)$, where $\neg \xi_R(\bar{v})$ is considered in NNF.
    \item If $\varphi(\bar{z}) := \circledast_{i \in I} \varphi_i(\bar{z})$ for $\circledast \in \{\bigwedge, \bigvee\}$, then $\Xi(\varphi) := \circledast_{i \in I} \Xi(\varphi_i)$. 
    \item If $\varphi(\bar{z})  := Q \bar{x} \varphi_1(\bar{z}, \bar{x})$ where $\varphi_1(\bar{z}, \bar{x}) := \circledast_{i \in I} \varphi'_i(\bar{z}, \bar{x})$ and $(Q, \circledast) \in $ $\{(\exists, \bigwedge),$ $ (\forall, \bigvee)\}$, then for $\bar{x} = (x_1, \ldots, x_r)$ for $r \ge 0$, we have:
    \begin{itemize}
        \item $\Xi(\varphi) := \exists \bar{x} (\bigwedge_{j \in [r]} \xi_U(x_j) \wedge \Xi(\varphi_1))$ if  $(Q, \circledast) = (\exists, \bigwedge)$.
        \item $\Xi(\varphi) := \forall \bar{x} ((\bigvee_{j \in [r]} \neg \xi_U(x_j)) $ $\vee \Xi(\varphi_1))$ if $(Q, \circledast) = (\forall, \bigvee)$, where $\neg \xi_U(x_j)$ is taken in NNF. (Observe that $\Xi(\varphi)$ here is equivalent to the formula  $\forall \bar{x} ((\bigwedge_{j \in [r]} \xi_U(x_j)) \rightarrow  \Xi(\varphi_1))$.)
    \end{itemize} 
\end{enumerate}
We can now state the following equivalence. (This is a special case of a more general result called the fundamental theorem of interpretations). For any $\tau$-structure $\str{A}$, a $(\tau, \sigma)$-interpretation $\Xi$, and an $\mc{L}$ sentence $\varphi$ over $\sigma$,
\begin{equation*}
\Xi(\str{A}) \models \varphi ~~\leftrightarrow~~ \str{A} \models \Xi(\varphi).
\end{equation*}
Applying this to the special setting of $\Xi$ being the definition of a quantifier-free sum-like binary  operation $\divideontimes$, we get for $\tau$-structures $\str{A}_1$ and $\str{A}_2$ and an $\mc{L}$ sentence $\varphi$ over $\tau$, that
\begin{equation}\label{thm:intp}
\str{A}_1 \divideontimes \str{A}_2 \models \varphi ~~\leftrightarrow~~ \str{A}_1 \anncupdot \str{A}_2 \models \Xi(\varphi).
\end{equation}
This equivalence will be useful for us in transferring results about $\anncupdot$ to similar results about $\divideontimes$.

\vspace{2pt} \noindent \tbf{3. Reduction sequences}: 
We now recall the notions of reduction sequences and models for these from the literature. We mention that reduction sequences as we present them below are an adaptation of the special case of 2-reduction sequences from~\cite{HHS15}, and the adaptation follows the ideas in~\cite{Gro08}.


Let $\mc{L}$ be a logic.  Given numbers $r \ge 0$ and $j \in [2]$, and index set $I$ and  an element $i \in I$, let $\psi_{i, j}$ be an $\mc{L}$ formula over a vocabulary $\tau$, whose free variables are contained in a (finite) sequence $\bar{x}_j$ of variables. We assume $\bar{x}_1$ and $\bar{x}_2$ to be disjoint. Let $\Delta_j(\bar{x}_j) = (\psi_{i, j})_{i \in I}$. Let $X_{i, j}$ be a propositional variable, $\mc{X} = \{ X_{i, j} \mid i \in I, j \in [2]\}$, and $\beta$ be an $\infty$-propositional formula (the terminology akin to that it~\cite{Kar59}) over the variables of $\mc{X}$. That is $\beta$ is a formula built from propositional variables using conjunctions and disjunctions of arbitrary arity. We call the triple  $D(\bar{x}_1, \bar{x}_2) = (\Delta_1(\bar{x}_1), \Delta_2(\bar{x}_2), \beta)$ an \emph{$\mc{L}$ reduction sequence over $\tau$}. If $\tau$ is clear from context, then we call $D(\bar{x}_1, \bar{x}_2)$ simply an $\mc{L}$ reduction sequence.

Let $\str{A}_1$ and $\str{A}_2$ be $\tau$-structures that are disjoint, and for $j \in [2]$, let $\bar{a}_j$ be a (finite) tuple of elements from $\str{A}_j$. We say that $(\str{A}_1, \str{A}_2, \bar{a}_1, \bar{a}_2)$ is a model of the $\mc{L}$ reduction sequence $D(\bar{x}_1, \bar{x}_2) = (\Delta_1(\bar{x}_1), \Delta_2(\bar{x}_2), \beta)$, denoted $(\str{A}_1, \str{A}_2, \bar{a}_1, \bar{a}_2) \models D(\bar{x}_1, \bar{x}_2)$, if $|\bar{a}_j| = |\bar{x}_j|$ for $j \in [2]$, and there exists an assignment $\zeta: \mc{X} \rightarrow \{0, 1\}$ such that $\zeta \models \beta$ and for $i \in I$ and $j \in [2]$, 
\[
\zeta(X_{i, j}) = 1  ~~~\leftrightarrow~~~ (\str{A}_j, \bar{a}_j) \models \psi_{i, j}(\bar{x}_j).
\]

Let $\divideontimes$ be a quantifier-free sum-like binary  operation on $\tau$-structures, and let $\ostar$ be one of the operations $\divideontimes$ or $\anncupdot$. Let $\mc{L}$ be a logic over $\tau$ as before, and let $\mc{L}_\ostar$ be a logic over $\tau$, respectively over $\anntau$, if $\ostar = \divideontimes$, respectively $\ostar = \anncupdot$. Given an $\mc{L}_\ostar$ formula $\varphi(\bar{x}_1, \bar{x}_2)$, we say that an $\mc{L}$ reduction sequence $D(\bar{x}_1, \bar{x}_2)$ is a \emph{Feferman-Vaught decomposition of $\varphi(\bar{x}_1, \bar{x}_2)$ over $\ostar$} if it holds that for any two  $\tau$-structures $\str{A}_1$ and $\str{A}_2$, if $\bar{a}_j$ is a $|\bar{x}_j|$-tuple from $\str{A}_j$ for  $j \in [2]$, then 
\[
(\str{A}_1 \ostar \str{A}_2, \bar{a}_1, \bar{a}_2) \models \varphi(\bar{x}_1, \bar{x}_2) ~~~\leftrightarrow~~~ (\str{A}_1, \str{A}_2, \bar{a}_1, \bar{a}_2) \models D(\bar{x}_1, \bar{x}_2)
\]
We also say that $\varphi$ \emph{factorizes} over $\divideontimes$ into its factors $\Delta_1(\bar{x}_1)$ and $\Delta_2(\bar{x}_2)$. 


Let $\ord$ denote the class of all ordinals and $\card$ the class of all cardinals. A function that will be important for us in this paper is $\mathsf{tower}: \ord \times \card \rightarrow \card$ which is the function defined as follows: $\tower{0, \kappa} = \kappa$, and $\tower{\lambda, \kappa} = 2^{\tower{\lambda-1, \kappa}}$. In the special case when $\lambda = n$ and $\kappa = k$ for $n, k \in \mathbb{N}$, the function $\tower{\lambda, \kappa}$ is an $n$-fold exponential function of $k$.


Finally, we abbreviate in the standard way the expressions `if and only if' as `iff', `with respect to' as `w.r.t.',  `respectively' as `resp.' and `without loss of generality' as `w.l.o.g'.





\section{{\fv} decompositions for $\tsinf{\kappa, \lambda}$ and $\tpinf{\kappa, \lambda}$}\label{section:gen-decomposition}

In this section, we show the existence of  Feferman-Vaught decompositions for formulae in $\tsinf{\kappa, \lambda}$ and $\tpinf{\kappa, \lambda}$ over quantifier-free sum-like operations. The following theorem is at the heart of our results in this and the next section.  To state the theorem, let $\infcard$ denote the class of all infinite cardinals. Define the function $\rho: \infcard  \times \ord \rightarrow \card$ as $\rho(\kappa, \lambda) = \tower{\lambda, \kappa}$ if $\kappa > \omega$, else  $\rho(\kappa, \lambda) = \omega$. Let $\hat{\rho}(\tsinf{\kappa, \lambda, r}[\mu]) = \tsinf{\rho(\kappa, \lambda), \lambda, r}[\mu]$ and $\hat{\rho}(\tpinf{\kappa, \lambda, r}[\mu]) = \tpinf{\rho(\kappa, \lambda), \lambda, r}[\mu]$, for each $r \in \mathbb{N}$; by extension let $\hat{\rho}(\tsinf{\kappa, \lambda}[\mu]) = \tsinf{\rho(\kappa, \lambda), \lambda}[\mu]$ and $\hat{\rho}(\tpinf{\kappa, \lambda}[\mu]) = \tpinf{\rho(\kappa, \lambda), \lambda}[\mu]$. Recall that for a vocabulary $\tau$, the vocabulary $\anntau = \tau \cupdot  \{P\}$ for a unary predicate $P$ (not in $\tau$) is the vocabulary of the annotated disjoint unions of $\tau$-structures.

\begin{theorem}\label{thm:gen-decomposition}
Let $\mc{L}$ be one of the logics $\tsiginf{\kappa, \lambda}[\mu]$ or $\tpinf{\kappa, \lambda}[\mu]$ for cardinals $\kappa, \mu$ and ordinal $\lambda$ such that $\kappa \ge \omega$ and $\mu, \lambda \ge 0$. Let $\tau$ be a vocabulary. Then for 
each $\mc{L}$ formula $\varphi(\bar{x}_1, \bar{x}_2)$ over $\anntau$, there is a $\rhohat(\mc{L})$ reduction sequence over $\tau$ that is a Feferman-Vaught decomposition of $\varphi(\bar{x}_1, \bar{x}_2)$ over the annotated disjoint union operation.
\end{theorem}

\begin{proof} 
We prove the theorem by showing a stronger  statement $\mc{P}(\lambda, \mc{L})$ as below. \\[5pt]
\noindent \vspace{3pt}
\begin{tabular}{p{0.15\textwidth}p{0.8\textwidth}}
$\mc{P}(\lambda, \mc{L}) ~~\equiv$ & For each $\mc{L}$ formula $\varphi(\bar{x}_1, \bar{x}_2)$ over $\anntau$, there is a $\hat{\rho}(\mc{L})$-reduction sequence $D(\bar{x}_1, \bar{x}_2) = (\Delta_1(\bar{x}_1), \Delta_2(\bar{x}_2), \beta)$ over $\tau$ that is a Feferman-Vaught decomposition of $\varphi(\bar{x}_1, \bar{x}_2)$ over the annotated disjoint union operation, and is such that (i) $\beta$ contains no negations, and (ii) if $\lambda > 0$ and $\varphi(\bar{x}_1, \bar{x}_2) \in \tsinf{\kappa, \lambda, r}$ ($\tpinf{\kappa, \lambda, r}$), then for $\eta < \rho(\kappa, \lambda)$, the formula $\beta$ is an $\eta$-ary disjunction (conjunction) of conjuncts (disjuncts) that are each a conjunction (disjunction) of exactly two positive literals, one a variable corresponding to a formula in $\Delta_1$ and the other a variable corresponding to a formula in $\Delta_2$.
\end{tabular}


\vspace{3pt}Our proof goes via showing $\mc{P}(\lambda, \tsiginf{\kappa, \lambda}[\mu])$ and $\mc{P}(\lambda, \tpiinf{\kappa, \lambda}[\mu])$ by simultaneous induction as $\lambda$ increases, for all $\kappa \ge \omega$ and $\mu \ge 0$. The analysis in the proof builds on the exposition in~\cite{Gro08}.


\vspace{3pt}\noindent \tbf{A. Base case}\label{thm:gen-decomposition:base-case}: The base case is when $\lambda = 0$. Note that in this case  $\tsiginf{\kappa, 0}[\mu] = \tpiinf{\kappa, 0}[\mu] = \tsiginf{\kappa, 0}[0] = \tpiinf{\kappa, 0}[0] = \tsiginf{\omega, 0} = \tpiinf{\omega, 0}$. We have the following subcases. It is easy to see in each case that the mentioned reduction sequence $D(\bar{x}_1, \bar{x}_2)$ 
is indeed a Feferman-Vaught decomposition of $\varphi(\bar{x}_1, \bar{x}_2)$. Observe that in each subcase, $D(\bar{x}_1, \bar{x}_2)$ is a $\tsiginf{\omega, 0}$ reduction sequence over $\tau$, and that the formula $\beta$ does not contain any negations.
\begin{enumerate}
    \item \label{atomic:non-annot} $\varphi(\bar{x}_1, \bar{x}_2) := A(\bar{z})$ where $A(\bar{z})$ is an atomic formula of the form $R(\bar{z})$ or $z_1 = z_2$ or the negations of these, for a predicate $R \in \tau$.
    \begin{enumerate}
        \item If $\bar{z}$ is a subtuple of $\bar{x}_1$, then the reduction sequence $D(\bar{x}_1, \bar{x}_2) = (\Delta_1(\bar{x}_1),$ $ \Delta_2(\bar{x}_2), \beta)$ is such that $\Delta_1(\bar{x}_1) = (A(\bar{z})), \Delta_2(\bar{x}_2) = (\true)$ and $\beta = X_{1, 1} \wedge X_{1, 2}$. 
        \item If $\bar{z}$ is a subtuple of $\bar{x}_2$, then the reduction sequence is $D(\bar{x}_1, \bar{x}_2)$ as above but where  $\Delta_1(\bar{x}_1) = (\true), \Delta_2(\bar{x}_2) = (A(\bar{z}))$ and $ \beta = X_{1, 1} \wedge X_{1, 2}$.
        \item If $\bar{z}$ is a neither a subtuple of $\bar{x}_1$ nor of $\bar{x}_2$, then the reduction sequence is $D(\bar{x}_1, \bar{x}_2)$ where $\Delta_1(\bar{x}_1) = \Delta_2(\bar{x}_2) = () ~( =$ the empty tuple), and $ \beta = \false$ if $A$ does not contain negation, and $\beta = \true$ if $A$ contains negation.
        \end{enumerate}
    \item \label{atomic:annot} $\varphi(\bar{x}_1, \bar{x}_2) := A(z)$ where $A(z)$ is the atomic formula $P(z)$ or its negation. Then the reduction sequence is $D(\bar{x}_1, \bar{x}_2) = (\Delta_1(\bar{x}_1), \Delta_2(\bar{x}_2), \beta)$ where $\Delta_1(\bar{x}_1) = \Delta_2(\bar{x}_2) = ()$ and 
    \begin{itemize}
        \item $\beta = \true$ if either $z$ is a variable of $\bar{x}_1$ and $A(z) = P(z)$, or $z$ is a variable of $\bar{x}_2$ and $A(z) = \neg P(z)$.
        \item $\beta = \false$ otherwise
    \end{itemize}
    \item \label{andor}
    $\varphi(\bar{x}_1, \bar{x}_2) :=  \varphi_1(\bar{x}_{1, 1}, \bar{x}_{2, 1}) \circledast \varphi_2(\bar{x}_{1, 2}, \bar{x}_{2, 2})$ where $\circledast \in \{\wedge, \vee\}$, $\varphi_k$ is quantifier-free, and $\bar{x}_{j, k}$ is a subtuple of $\bar{x}_j$, for $j, k \in [2]$. Assume that there exist $\tsiginf{\kappa, 0}$ reduction sequences $D_k(\bar{x}_{1, k}, \bar{x}_{2, k}) = (\Delta^k_1(\bar{x}_{1, k}), \Delta^k_2(\bar{x}_{2, k}), \beta_k)$ that witness $\mc{P}(0, \tsiginf{\kappa, 0})$ for $\varphi_k(\bar{x}_{1, k}, \bar{x}_{2, k})$, for $k \in [2]$. Then the desired reduction sequence for $\varphi(\bar{x}_1, \bar{x}_2)$ is $D(\bar{x}_1, \bar{x}_2) = (\Delta_1(\bar{x}_1), \Delta_2(\bar{x}_2), \beta)$ where $\Delta_j(\bar{x}_j) = \Delta^1_j(\bar{x}_{j, 1}) \cdot \Delta^2_j(\bar{x}_{j, 2})$ for $j \in [2]$, and $\beta = \beta_1 \circledast \beta_2$.  Here $\cdot$ denotes concatenation of tuples.

\end{enumerate}
\noindent \tbf{B. Induction:} Assume as induction hypothesis, that $\mc{P}(\lambda', \mc{L}')$ holds for $\mc{L}'$ that is one of the logics $\tsiginf{\kappa', \lambda'}[\mu']$ or $\tpiinf{\kappa', \lambda'}[\mu']$ over $\tau$, for all $\kappa', \mu', \lambda'$ such that $\kappa' \ge \omega, \mu' \ge 0$ and $\lambda' < \lambda$ where $\lambda > 0$. We show below that $\mc{P}(\lambda, \mc{L})$ holds for the case when $\mc{L} = \tsiginf{\kappa, \lambda}[\mu]$ for an arbitrary $\kappa, \mu$ such that $\kappa \ge \omega$ and $\mu \ge 0$. The reasoning when $\mc{L} = \tpiinf{\kappa, \lambda}[\mu]$ can be similarly done (by considering disjunctions in place of conjunctions and vice-versa, and universal quantifiers in place of existential quantifiers and vice-versa) to complete the induction. We recall from Section~\ref{section:prelims} that $\tsiginf{\kappa, \lambda}[\mu] = \bigcup_{r \in \mathbb{N}} \tsiginf{\kappa, \lambda, r}[\mu]$. Our proof below goes via showing $\mc{P}(\lambda, \tsinf{\kappa, \lambda, r}[\mu])$ by a nested induction on $r$.


\vspace{2pt}\noindent \tbf{Nested base case:} The base case is when $r = 0$. Then $\varphi(\bar{x}_1, \bar{x}_2)$ is given by $\varphi(\bar{x}_1, \bar{x}_2) := \bigwedge_{i \in I} \varphi_i(\bar{x}_{1, i}, \bar{x}_{2, i})$ where $I$ is an index set of cardinality $< \kappa$, $\varphi_i$ is a formula of $\tpiinf{\kappa, \lambda_i}[\mu]$ over $\anntau$ for some $\lambda_i < \lambda$, and $\bar{x}_{j, i}$ is a subtuple of $\bar{x}_j$ for $j \in [2]$ and $i \in I$. From the (outer) induction hypothesis above, let $D_i(\bar{x}_{1, i}, \bar{x}_{2, i}) = (\Delta^i_1(\bar{x}_{1, i}), \Delta^i_2(\bar{x}_{2, i}), \beta_i)$ be the $\tpiinf{\rho(\kappa, \lambda_i), \lambda_i}[\mu]$ reduction sequence over $\tau$ that witnesses $\mc{P}(\lambda_i,$ $ \tpiinf{\kappa, \lambda_i}[\mu])$ for $\varphi_i(\bar{x}_{1, i}, \bar{x}_{2, i})$, for $i \in I$. We have two cases as below:

\begin{enumerate}
    \item \label{beta-i-form} $\lambda > 1$: Then $\beta_i$ is of the form $\bigwedge_{j \in J_i} (X^{(i, j)}_1 \vee X^{(i. j)}_2)$ where $J_i$ is an index set of cardinality $< \rho(\kappa, \lambda_i)$, and if $X^{(i, j)}_l$ corresponds to the formula $\psi^{(i, j)}_l$, then $\Delta^i_l = (\psi^{(i, j)}_l)_{j \in J_i}$ for $l \in [2]$. 

    \item $\lambda = 1$: In this case, $\beta_i$ is a finite propositional formula. Writing $\beta_i$ as an AND of ORs, we get $\beta_i \leftrightarrow \bigwedge_{j \in J_i} ((\bigvee_{l = 1}^{l = j_1} Z^{(i, j)}_{1, l}) \vee (\bigvee_{l = 1}^{l = j_2} Z^{(i, j)}_{2, l}))$ where $|J_i|, j_1, j_2 < \omega$, the numbers $j_1, j_2$ are non-zero w.l.o.g., and for $k \in [2]$, $Z^{(i, j)}_{k, l}$ corresponds to the quantifier-free FO formula $\chi^{(i, j)}_{k, l}$ and  $\Delta^i_k = (\chi^{(i, j)}_{k, l})_{j \in J_i, l \in [j_k]}$ for $k \in [2]$. Consider the $\hat{\rho}(\tsinf{\kappa, 0})$ reduction sequence  $\ul{D}^{i}(\bar{x}_{1, i}, \bar{x}_{2, i}) = (\ul{\Delta}^i_1(\bar{x}_{1, i}), $ $ \ul{\Delta}^i_2(\bar{x}_{2, i}),$ $ \beta'_i)$ such that $\ul{\Delta}^i_k(\bar{x}_{k, i}) =  (\psi^{(i, j)}_k)_{j \in J_i}$  where $\psi^{(i, j)}_k := \bigvee_{l = 1}^{l = j_k} \chi^{(i, j)}_{k, l}$ for $k \in [2]$ and $\beta'_i = \bigwedge_{j \in J_i} (X^{(i, j)}_1 \vee X^{(i, j)}_2)$ where $X^{(i, j)}_k$ is a new propositional variable that corresponds to $\psi^{(i, j)}_k$ for $k \in [2]$. It is easy to see that $\ul{D}^i(\bar{x}_{1, i}, \bar{x}_{2, i})$ is ``equivalent" to $D^i(\bar{x}_{1, i}, \bar{x}_{2, i})$, in that $\ul{D}^i(\bar{x}_{1, i}, \bar{x}_{2, i})$ is also a Feferman-Vaught decomposition of $\varphi_i(\bar{x}_{1, i}, \bar{x}_{2, i})$. 
\end{enumerate}
In either case therefore, we can w.l.o.g. consider $\beta_i$ to be of the form as stated in case (\ref{beta-i-form}) above.

Let $J =  \{ (i, j) \mid i \in I, j \in J_i\}$. Consider the formula $\beta' = \bigwedge_{i \in I} \beta_i$. Writing this formula as an OR of ANDs, we have that

    \begin{equation}\label{basecase:orofands}
    \begin{aligned}
    \beta' \leftrightarrow \beta''  := \bigvee_{f \in \{1, 2\}^{J}} C_f ~~~~\mbox{where}~~~~ 
     C_f := \bigwedge_{k \in S_{f, 1}} X^{k}_1 \wedge \bigwedge_{k \in S_{f, 2}} X^{k}_2
    \end{aligned}        
    \end{equation}
    
    Above $\{1,2\}^{J}$ denotes the set of all functions $f: J \rightarrow \{1, 2\}$, the set $J$ is partitioned into $S_{f, 1}$ and $S_{f, 2}$ (allowing empty parts), where $S_{f, l} = \{p \in J \mid f(p) = l\}$ for $l \in [2]$. We now define the formulae $\xi_{f, l}(\bar{x}_l)$ for $f \in \{1, 2\}^J$ and $l \in [2]$ as below.
    \begin{align}\label{basecase:newformulae}
        \xi_{f, l}(\bar{x}_l) & := \bigwedge\limits_{\substack{k \in S_{f, l}\\k = (i, j)}} \psi^k_l(\bar{x}_{l, i})
    \end{align}
    In the event that $S_{f, l} = \emptyset$, we put $\xi_{f, l}(\bar{x}_l) := \true$. Let $Y_{f, l}$ be a new propositional variable for $f \in \{1, 2\}^J$ and $l \in [2]$. Consider the reduction sequence $D(\bar{x}_1, \bar{x}_2) = (\Delta_1(\bar{x}_1), \Delta_2(\bar{x}_2), \beta)$ where for $l \in [2]$
    \begin{align}\label{basecase:redseq}
        \Delta_l(\bar{x}_l) & = (\xi_{f, l}(\bar{x}_l))_{f \in \{1, 2\}^J} ~~~~ ;~~~~ \beta := \bigvee_{f \in \{1, 2\}^J} (Y_{f, 1} \wedge Y_{f, 2})
    \end{align}

    We claim that $D(\bar{x}_1, \bar{x}_2)$ witnesses $\mc{P}(\lambda, \tsinf{\kappa, \lambda, 0}[\mu])$ for $\varphi(\bar{x}_1, \bar{x}_2)$. 

    \begin{enumerate}
        \item Firstly, $|J| = |I| \cdot \max\{|J_i| \mid i \in I\} < \kappa \cdot \max\{|J_i| \mid i \in I\}\} < \max\{\rho(\kappa, \lambda_i) \mid i \in I\}$. So that 
        \begin{equation}\label{basecase:beta-calculations}
        |\{1, 2\}^J| = 2^{|J|} < 
        \left\{ 
            \begin{array}{ll}
                \begin{aligned}
                \phantom{=} & ~~\omega    
                \end{aligned} & ~~~~\mbox{if}~\kappa = \omega  \\ 
                \begin{aligned}
                \phantom{=} & \phantom{\omega}    
                \end{aligned} & \\                \begin{aligned}
                    & ~\max\{2^{\rho(\kappa, \lambda_i)} \mid i \in I\} \\
                    = & ~\max\{2^{\tower{\lambda_i, \kappa}} \mid i \in I\} \\
                    = & ~\max\{\tower{\lambda_i+1, \kappa} \mid i \in I\}\\
                    \leq & ~\tower{\lambda, \kappa}\\
                \end{aligned} & ~~~~\mbox{if}~\kappa > \omega
                \end{array}
            \right.
        \end{equation}
        Then $\beta$ is indeed of the form required by $\mc{P}(\lambda, \tsinf{\kappa, \lambda, 0}[\mu])$ for $\varphi(\bar{x}_1, \bar{x}_2)$.
        \item The formula $\psi^k_l(\bar{x}_{l, i})$  in~(\ref{basecase:newformulae}) belongs to $\tpinf{\rho(\kappa, \lambda_i), \lambda_i}[\mu]$ over $\tau$ by induction hypothesis. And $\xi_{f, l}(\bar{x}_l)$ is a $\delta$-ary conjunction of the $\psi^k_l$s (for $k$ ranging over $S_{f, l}$), where $\delta = |S_{f, l}| \leq |J| < \max\{\rho(\kappa, \lambda_i) \mid i \in I\} \leq \rho(\kappa, \lambda)$. Hence $\xi_{f, l}(\bar{x}_l)$ is in $\tsinf{\rho(\kappa, \lambda), \lambda, 0}[\mu]$ over $\tau$. Then $D(\bar{x}_1, \bar{x}_2)$ is a $\hat{\rho}(\tsinf{\kappa, \lambda, 0}[\mu])$-reduction sequence over $\tau$.
        \item Finally, the reduction sequence $D(\bar{x}_1, \bar{x}_2)$ is a Feferman-Vaught decomposition for $\varphi(\bar{x}_1, \bar{x}_2)$ as seen via the following equivalences. Below, the third equivalence is by the induction hypothesis; $\bar{a}_{l, i}$ is the sub-tuple of $\bar{a}_l$ corresponding to $\bar{x}_{l, i}$; $\mc{X}_i = \{ X^{(i, j)}_l \mid j \in J_i, l \in [2]\}$ for $i \in I$; and $\mc{Y} = $ $\{ Y_{f, l} \mid$ $ f \in \{1, 2\}^J, l \in [2]\}$.
        
        \[\def\arraystretch{1.3}
        \begin{array}{ll}
            & (\str{A}_1 \anncupdot \str{A}_2, \bar{a}_1, \bar{a}_2) \models \varphi(\bar{x}_1, \bar{x}_2) \\
            
            \leftrightarrow & (\str{A}_1 \anncupdot \str{A}_2, \bar{a}_1, \bar{a}_2) \models \bigwedge_{i \in I} \varphi_i(\bar{x}_{1, i}, \bar{x}_{2, i}) \\
            
            \leftrightarrow & \bigwedge_{i \in I} (\str{A}_1 \anncupdot \str{A}_2, \bar{a}_{1, i}, \bar{a}_{2, i}) \models \varphi_i(\bar{x}_{1, i}, \bar{x}_{2, i})\\
     
            \leftrightarrow & \bigwedge_{i \in I} (\str{A}_1, \str{A}_2, \bar{a}_{1, i}, \bar{a}_{2, i}) \models D_i(\bar{x}_{1, i}, \bar{x}_{2, i})\\ 
     
            \leftrightarrow & \mbox{For all}~i \in I,~\mbox{there exists}~\zeta_i:\mc{X}_i \rightarrow \{0, 1\} ~\mbox{s.t.}~ \zeta_i \models \beta_i~\mbox{and}\\
            
            & \zeta_i(X^{(i, j)}_l) = 1 ~\mbox{iff}~ (\str{A}_l, \bar{a}_{l, i})  \models \psi^{(i, j)}_l(\bar{x}_{l, i})~~~~~\mbox{for}~ j\in J_i~\mbox{and}~l \in[2]\\
            
            \leftrightarrow & \mbox{For all}~i \in I~\mbox{there exists}~\zeta_i:\mc{X}_i \rightarrow \{0, 1\}~\mbox{s.t. for all}~j \in J_i,~\mbox{there exists}\\
            
            &~l\in[2] ~\mbox{s.t.}~\zeta_i \models X^{(i, j)}_l~\mbox{and}\\         
        
            & \zeta_i(X^{(i, j)}_l) = 1 ~\mbox{iff}~ (\str{A}_l, \bar{a}_{l, i})  \models \psi^{(i, j)}_l(\bar{x}_{l, i})\\
            
        
            \leftrightarrow &\mbox{There exists}~\zeta:\mc{Y} \rightarrow \{0, 1\} ~\mbox{s.t.}~ \zeta \models Y_{f, 1} \wedge Y_{f, 2}~\mbox{for some}~f \in \{1, 2\}^J~\mbox{and}\\
            
            & \zeta(Y_{f, l}) = 1 ~\mbox{iff}~ (\str{A}_l, \bar{a}_l)  \models \xi_{f, l}(\bar{x}_l) ~~~\mbox{for}~l \in [2]\\

            \leftrightarrow &\mbox{There exists}~~\zeta:\mc{Y} \rightarrow \{0, 1\} ~\mbox{s.t.}~ \zeta \models \beta~\mbox{and}\\ 
    
            & \zeta(Y_{f, l}) = 1 ~\mbox{iff}~ (\str{A}_l, \bar{a}_l)  \models \xi_{f, l}(\bar{x}_l) ~~~\mbox{for}~f \in \{1, 2\}^J~\mbox{and}~l \in[2]\\
    
            \leftrightarrow & (\str{A}_1, \str{A}_2, \bar{a}_1, \bar{a}_2) \models D(\bar{x}_1, \bar{x}_2)\\
    \end{array}
    \]
    \end{enumerate}
    This establishes the nested base case.
    
\vspace{2pt}\noindent \tbf{Nested induction:} Assume as the nested induction hypothesis that $\mc{P}(\lambda,$ $ \tsinf{\kappa, \lambda, r}[\mu])$ holds for $r = r_0 \ge 0$. Consider a formula $\varphi(\bar{x}_1, \bar{x}_2)$ of $\tsiginf{\kappa, \lambda,  r_0+1}[\mu]$ over $\anntau$.  The formula has the form $\varphi(\bar{x}_1, \bar{x}_2) := \exists z \varphi_1(\bar{x}_1, \bar{x}_2, z)$ where $\varphi_1$ is a formula of $\tsiginf{\kappa, \lambda, r_0}[\mu]$ over $\anntau$. We observe that the free variables of $\varphi_1$ can be seen as being amongst the tuple $\bar{y}_{1} \cdot \bar{y}_{2}$ where either $\bar{y}_1 = \bar{x}_1 \cdot z$ and $\bar{y}_2 = \bar{x}_2$, or $\bar{y}_1 = \bar{x}_1$ and $\bar{y}_2 = \bar{x}_2 \cdot z$. Corresponding to each of these views, we have by the nested induction hypothesis that there exist $\tsiginf{\rho(\kappa, \lambda), \lambda, r_0}[\mu]$ reduction sequences $D_1(\bar{x}_1 \cdot z, \bar{x}_2) = (\Delta^1_1(\bar{x}_1 \cdot z), \Delta^1_2(\bar{x}_2), \beta_1)$ and $D_2(\bar{x}_1, \bar{x}_2 \cdot z) = (\Delta^2_1(\bar{x}_1), \Delta^2_2(\bar{x}_2 \cdot z), \beta_2)$ over $\tau$ witnessing $\mc{P}(\lambda, \tsinf{\kappa, \lambda, r_0}[\mu])$ resp. for $\varphi_1(\bar{x}_1 \cdot z, \bar{x}_2)$ and $\varphi_1(\bar{x}_1, \bar{x}_2 \cdot z)$.

Let $\Delta^1_1(\bar{x}_1 \cdot z) = (\psi^{(i, 1)}_1(\bar{x}_1 \cdot z))_{i \in I_1}, \Delta^1_2 = (\psi^{(i, 1)}_2(\bar{x}_2))_{i \in I_1}, \Delta^2_1 = (\psi^{(i, 2)}_1(\bar{x}_1))_{i \in I_2},$ and $\Delta^2_2 = (\psi^{(i, 2)}_2(\bar{x}_2 \cdot z))_{i \in I_2}$. Let $\beta_j := \bigvee_{i \in I_j} (X^{(i, j)}_1 \wedge X^{(i, j)}_2)$ -- observe that by the nested induction hypothesis this is the form of $\beta_j$ -- for $j \in [2]$, where $X^{(i, j)}_l$ corresponds to the formula $\psi^{(i, j)}_l$ for $i \in I_j, l \in [2]$, and $|I_j| < \rho(\kappa, \lambda)$.

We now define the formulae $\xi^{(i, j)}_l(\bar{x}_j)$ for $j, l \in [2]$ and $i \in I_j$ as below.
\begin{equation}\label{induction:newformulae}
    \begin{split}
        \xi^{(i, 1)}_1(\bar{x}_1) &  :=  \exists z \psi^{(i, 1)}_1(\bar{x}_1, z)\\
        \xi^{(i, 2)}_1(\bar{x}_1) &  :=   \psi^{(i, 2)}_1(\bar{x}_1) 
    \end{split}
    \quad  \quad 
    \begin{split}
        \xi^{(i, 1)}_2(\bar{x}_2) &  :=  \psi^{(i, 1)}_2(\bar{x}_2)\\
        \xi^{(i, 2)}_2(\bar{x}_2) &  :=   \exists z \psi^{(i, 2)}_2(\bar{x}_2, z) 
    \end{split}
\end{equation}
Let $Y^{(i, j)}_l$ be a new propositional variable for $j, l \in [2]$ and $i \in I_j$. Consider the reduction sequence $D(\bar{x}_1, \bar{x}_2) = (\Delta_1(\bar{x}_1), \Delta_2(\bar{x}_2), \beta)$ where for $l \in [2]$ 
\begin{align}\label{induction:redseq}
    \Delta_l(\bar{x}_l) & = (\xi^{(i, 1)}_l)_{i \in I_1} \cdot (\xi^{(i, 2)}_l)_{i \in I_2} ~~~~ ;~~~~ \beta := \bigvee_{j \in [2]} \bigvee_{i \in I_j} (Y^{(i, j)}_1 \wedge Y^{(i, j)}_2)
\end{align}

We claim that $D(\bar{x}_1, \bar{x}_2)$ witnesses $\mc{P}(\lambda, \tsinf{\kappa, \lambda, r_0+1}[\mu])$ for $\varphi(\bar{x}_1, \bar{x}_2)$. 

\begin{enumerate}
    \item By the nested induction hypothesis, $|I_j| < \rho(\kappa, \lambda)$ for $j \in [2]$. Then $|I_1| + |I_2| < \rho(\kappa, \lambda)$ whereby $\beta$ is indeed as required by $\mc{P}(\lambda, \tsinf{\kappa, \lambda, r_0+1}[\mu])$ for $\varphi(\bar{x}_1, \bar{x}_2)$.
    \item The formula $\varphi_1$ is in $\tsiginf{\kappa, \lambda, r_0}[\mu]$ over $\anntau$; so by the nested induction hypothesis, $\psi^{(i, j)}_l$  is a $\tsiginf{\rho(\kappa, \lambda), \lambda, r_0}[\mu]$ formula over $\tau$. Then $\xi^{(i, j)}_l$ is a formula of $\tsiginf{\rho(\kappa, \lambda), \lambda, r_0+1}[\mu] = \hat{\rho}(\tsinf{\kappa, \lambda, r_0+1}[\mu])$ over $\tau$. Hence $D(\bar{x}_1, \bar{x}_2)$ is  a $\hat{\rho}(\tsinf{\kappa, \lambda, r_0+1}[\mu])$ reduction sequence over $\tau$.
    \item The reduction sequence $D(\bar{x}_1, \bar{x}_2)$ is a Feferman-Vaught decomposition for $\varphi(\bar{x}_1, \bar{x}_2)$, which we show using the equivalences below. 
    Below, $b$ is an element of $\str{A}_1 \anncupdot \str{A}_2$; the third equivalence is by the induction hypothesis; $\mc{X}_j = \{ X^{(i, j)}_l \mid i \in I_j, l \in [2]\}$ for $j \in [2]$; and $\mc{Y} = \{ Y^{(i, j)}_l \mid j, l \in [2], i \in I_j\}$.
    \[\def\arraystretch{1.3}
    \begin{array}{ll}
        & (\str{A}_1 \anncupdot \str{A}_2, \bar{a}_1, \bar{a}_2) \models \varphi(\bar{x}_1, \bar{x}_2) \\
        
        \leftrightarrow & (\str{A}_1 \anncupdot \str{A}_2, \bar{a}_1, \bar{a}_2) \models \exists z \varphi_1(\bar{x}_1, \bar{x}_2, z) \\

        \leftrightarrow &  (\str{A}_1 \anncupdot \str{A}_2, \bar{a}_1 \cdot b, \bar{a}_2) \models \varphi_1(\bar{x}_1 \cdot z, \bar{x}_2) \bigvee  (\str{A}_1 \anncupdot \str{A}_2, \bar{a}_1, \bar{a}_2 \cdot b) \models \varphi_1(\bar{x}_1, \bar{x}_2  \cdot z)\\
     
        \leftrightarrow & (\str{A}_1, \str{A}_2, \bar{a}_1 \cdot b, \bar{a}_2) \models D_1(\bar{x}_1 \cdot z, \bar{x}_2) \bigvee (\str{A}_1, \str{A}_2, \bar{a}_1, \bar{a}_2 \cdot b) \models D_2(\bar{x}_1, \bar{x}_2  \cdot z)\\
     
        \leftrightarrow & \mbox{For some}~j\in [2], ~\mbox{there exists}~\zeta_j:\mc{X}_j \rightarrow \{0, 1\} ~\mbox{s.t.}~ \zeta_j \models \beta_j~\mbox{and}\\
        
        & \mbox{if}~ j = 1,~\mbox{then}\\ 
        
        & \begin{array}{ll}
            \zeta_1(X^{(i, 1)}_1) = 1 ~\mbox{iff}~ (\str{A}_1, \bar{a}_1 \cdot b)  \models \psi^{(i, 1)}_1(\bar{x}_1 \cdot z)  &~~~~i \in I_1\\
             
            \zeta_1(X^{(i, 1)}_2) = 1 ~\mbox{iff}~ (\str{A}_2, \bar{a}_2)  \models \psi^{(i, 1)}_2(\bar{x}_2)  &~~~~i \in I_1
        \end{array}\\
        
        & \mbox{else}\\
        
        & \begin{array}{ll}
            \zeta_2(X^{(i, 2)}_1) = 1 ~\mbox{iff}~ (\str{A}_1, \bar{a}_1)  \models \psi^{(i, 2)}_1(\bar{x}_1)  &~~~~i \in I_2\\

            \zeta_2(X^{(i, 2)}_2) = 1 ~\mbox{iff}~ (\str{A}_2, \bar{a}_2 \cdot b)  \models \psi^{(i, 2)}_2(\bar{x}_2 \cdot z)  &~~~~i \in I_2
        \end{array}\\     

    
              \leftrightarrow & \mbox{For some}~j\in [2], ~\mbox{there exists}~\zeta_j:\mc{X}_j \rightarrow \{0, 1\}~\mbox{s.t. for some}~i \in I_j,  \\
        & \zeta_j \models (X^{(i, j)}_1 \wedge X^{(i, j)}_2)~\mbox{and}\\
        
        & \mbox{if}~ j = 1,~\mbox{then}\\ 
         
        & \begin{array}{ll}
            \zeta_1(X^{(i, 1)}_1) = 1 ~\mbox{iff}~ (\str{A}_1, \bar{a}_1 \cdot b)  \models \psi^{(i, 1)}_1(\bar{x}_1 \cdot z)  &~~~~i \in I_1\\
             
            \zeta_1(X^{(i, 1)}_2) = 1 ~\mbox{iff}~ (\str{A}_2, \bar{a}_2)  \models \psi^{(i, 1)}_2(\bar{x}_2)  &~~~~i \in I_1
         \end{array}\\
        
        & \mbox{else}~\\
        
        & \begin{array}{ll}
            \zeta_1(X^{(i, 2)}_1) = 1 ~\mbox{iff}~ (\str{A}_1, \bar{a}_1)  \models \psi^{(i, 2)}_1(\bar{x}_1)  &~~~~i \in I_2\\

            \zeta_1(X^{(i, 2)}_2) = 1 ~\mbox{iff}~ (\str{A}_2, \bar{a}_2 \cdot b)  \models \psi^{(i, 2)}_2(\bar{x}_2 \cdot z)  &~~~~i \in I_2
        \end{array}\\     
    \end{array}
    \]
    \[\def\arraystretch{1.3}
    \begin{array}{ll}
        \leftrightarrow &\mbox{There exists}~\zeta:\mc{Y} \rightarrow \{0, 1\} ~\mbox{s.t.}~ \zeta \models (Y^{(i, j)}_1 \wedge Y^{(i, j)}_2)~\mbox{for some}~j \in [2]~\mbox{and}\\
        
        &i\in I_j~\mbox{and}\\ 
    
        & \zeta(Y^{(i, j)}_l) = 1 ~\mbox{iff}~ (\str{A}_l, \bar{a}_l)  \models \xi^{(i, j)}_l(\bar{x}_l) ~~~~~\mbox{for}~l \in [2]\\
        
        \leftrightarrow &\mbox{There exists}~\zeta:\mc{Y} \rightarrow \{0, 1\} ~\mbox{s.t.}~ \zeta \models \beta~\mbox{and}\\ 
    
        & \zeta(Y^{(i, j)}_l) = 1 ~\mbox{iff}~ (\str{A}_l, \bar{a}_l)  \models \xi^{(i, j)}_l(\bar{x}_l) ~~~~~\mbox{for}~i \in I_j~\mbox{and}~j, l \in [2]\\

        \leftrightarrow & (\str{A}_1, \str{A}_2, \bar{a}_1, \bar{a}_2) \models D(\bar{x}_1, \bar{x}_2)\\
    \end{array}
    \]
\end{enumerate}
This completes the nested induction, and hence the outer induction and the proof.
\end{proof}

\begin{corollary}\label{cor:gen-infty-decomposition}
Let $\mc{L}$ be one of the logics $\tsiginf{\infty, \lambda}[\mu]$ or $\tpiinf{\infty, \lambda}[\mu]$ for $\lambda, \mu \ge 0$. Let $\tau$ be a vocabulary. Then for each $\mc{L}$ formula $\varphi(\bar{x}_1, \bar{x}_2)$ over $\anntau$, there is an $\mc{L}$ reduction sequence over $\tau$ that is a Feferman-Vaught decomposition of $\varphi(\bar{x}_1, \bar{x}_2)$ over the annotated disjoint union operation.
\end{corollary}
\begin{proof}
Since $\varphi(\bar{x}_1, \bar{x}_2)$ is an $\mc{L}$ formula, it is a formula of the logic $\mc{L}_\kappa$ for some $\kappa \ge \omega$, where $\mc{L}_\kappa$ is the logic $\tsinf{\kappa, \lambda}[\mu]$  if $\mc{L}$ is $\tsinf{\infty, \lambda}[\mu]$, else $\mc{L}_\kappa$ is the logic $\tpinf{\kappa, \lambda}[\mu]$. By  Theorem~\ref{thm:gen-decomposition}, there is a $\rhohat(\mc{L}_\kappa)$ reduction sequence $D(\bar{x}_1, \bar{x}_2)$ over $\tau$ that is a Feferman-Vaught decomposition of $\varphi(\bar{x}_1, \bar{x}_2)$ over  annotated disjoint union. Since $\rhohat(\mc{L}_\kappa) = \mc{L}_{\rho(\kappa, \lambda)} \subseteq \mc{L}$, we have that $D(\bar{x}_1, \bar{x}_2)$ is also an $\mc{L}$ reduction sequence over $\tau$.
\end{proof}

\subsection{Decompositions over definable operations on structures}\label{subsection:genops}

We now consider quantifier-free sum-like operations on structures as defined in Section~\ref{section:prelims}, and show that these admit Feferman-Vaught decompositions for $\tsinf{\kappa, \lambda}$ and
$\tpinf{\kappa, \lambda}$.


\begin{theorem}\label{thm:gen-genop-decomposition}
Let $\kappa, \mu$ be cardinals and $\lambda$ be an  ordinal such that $\kappa \ge \omega$ and $\lambda, \mu \ge 0$. Let $\tau$ be a vocabulary and $\divideontimes$ be a quantifier-free sum-like binary operation on $\tau$-structures. Then the following are true: 
\begin{enumerate}
    \item If $\mc{L}$ is one of the logics $\tsinf{\kappa, \lambda}[\mu]$ or $\tpinf{\kappa, \lambda}[\mu]$ over $\tau$, then for every $\mc{L}$ sentence $\varphi$, there is a $\rhohat(\mc{L})$ reduction sequence that is a Feferman-Vaught decomposition of $\varphi$ over $\divideontimes$.\label{thm:gen-genop-decomposition:1} 
    \item If $\mc{L}$ is one of the logics $\tsinf{\infty, \lambda}[\mu]$ or $\tpinf{\infty, \lambda}[\mu]$ over $\tau$, then for every $\mc{L}$ sentence $\varphi$, there is an $\mc{L}$ reduction sequence that is a Feferman-Vaught decomposition of $\varphi$ over $\divideontimes$.\label{thm:gen-genop-decomposition:2}
\end{enumerate}
\end{theorem}
\begin{proof}
We show statement~(\ref{thm:gen-genop-decomposition:1}) above; statement~\ref{thm:gen-genop-decomposition:2} can be shown analogously using Corollary~\ref{cor:gen-infty-decomposition}. Let $\Xi$ be a quantifier-free definition of $\divideontimes$. Consider the $\mc{L}$ sentence $\psi := \Xi(\varphi)$ as defined in Section~\ref{section:prelims}. Let $D$ be the $\rhohat(\mc{L})$ reduction sequence for $\psi$ as given by Theorem~\ref{thm:gen-decomposition}, so $D$ is a Feferman-Vaught decomposition of $\psi$ over the annotated disjoint union operation. The following equivalences show that $D$ is also a Feferman-Vaught decomposition of $\varphi$ over $\divideontimes$. Let $\str{A}_1, \str{A}_2$ be $\tau$-structures.

\[\def\arraystretch{1.3}
\begin{array}{lll}
    & \str{A}_1 \divideontimes \str{A}_2 \models \varphi & \\
    \leftrightarrow & \str{A}_1 \anncupdot \str{A}_2 \models \Xi(\varphi) & ~~~(\mbox{by (\ref{thm:intp})})\\
    \leftrightarrow & \str{A}_1 \anncupdot \str{A}_2 \models \psi &~~~(\mbox{since}~\psi := \Xi(\varphi))\\
    \leftrightarrow & (\str{A}_1, \str{A}_2) \models D &
\end{array}
\]
\end{proof}

\begin{corollary}\label{cor:gen-composition}
Let $\mc{L}$ be one of the logics $\tsinf{\infty, \lambda}[\mu]$ or $\tpinf{\infty, \lambda}[\mu]$ over a vocabulary $\tau$, for $\lambda, \mu \ge 0$. Given (arbitrary) $\tau$-structures $\str{A}_1$ and $\str{A}_2$, and a quantifier-free sum-like binary operation $\divideontimes$ on $\tau$-structures, the $\mc{L}$ theory of $\str{A}_1 \divideontimes \str{A}_2$ is determined by the $\mc{L}$ theories of $\str{A}_1$ and $\str{A}_2$. 
\end{corollary}
\begin{proof}
Let $\str{A}_1', \str{A}_2'$ be $\tau$-structures such that $\str{A}_1 \equivl \str{A}_1'$ and $\str{A}_2 \equivl \str{A}_2'$ where $\equivl$ denotes indistinguishability with respect to all $\mc{L}$ sentences. Let $\varphi$ be an $\mc{L}$ sentence. We show the following to complete the proof. 
\begin{equation}
    \str{A}_1 \divideontimes \str{A}_2 \models \varphi ~~~\leftrightarrow~~~  \str{A}_1' \divideontimes \str{A}_2' \models \varphi \label{eqn:inf-comp:0}
\end{equation}
Towards showing (\ref{eqn:inf-comp:0}), let $D = (\Delta_1, \Delta_2, \beta)$ be the $\mc{L}$ reduction sequence for $\varphi$ over $\divideontimes$ as given by Theorem~\ref{thm:gen-genop-decomposition}(\ref{thm:gen-genop-decomposition:2}). Let $\psi_{i, j}$ for $i \in I, j \in [2]$ for an index set $I$ be $\mc{L}$ sentences such that $\Delta_j = (\psi_{i, j})_{i \in I}$. Let $X_{i, j}$ be propositional variables such that $\beta$ is an $\infty$-propositional formula over $\mc{X}   = \{X_{i, j} \mid i\in I, j \in [2]\}$. Then there exist assignments $\zeta, \zeta': \mc{X} \rightarrow \{0, 1\}$ such that for $i\in I$ and $j \in [2]$,
\begin{align}
\zeta(X_{i, j}) = 1 \leftrightarrow \str{A}_j \models \psi_{i, j} ~~&\text{and}~~ 
\zeta'(X_{i, j}) = 1 \leftrightarrow \str{A}'_j \models \psi_{i, j}\label{eqn:inf-comp:1}\\ 
\str{A}_1 \divideontimes \str{A}_2  \models \varphi \leftrightarrow  \zeta \models \beta ~~&\text{and}~~ \str{A}_1' \divideontimes \str{A}_2'  \models \varphi \leftrightarrow  \zeta' \models \beta \label{eqn:inf-comp:2} 
\end{align}
        
Since $\str{A}_j \equivl \str{A}_j'$, it follows that $\str{A}_j \models \psi_{i, j}$ iff $\str{A}_j' \models \psi_{i, j}$; whereby $\zeta(X_{i, j}) = 1$ iff $\zeta'(X_{i, j}) = 1$ from (\ref{eqn:inf-comp:1}). Then $\zeta = \zeta'$, so by (\ref{eqn:inf-comp:2}), we indeed have (\ref{eqn:inf-comp:0}).
\end{proof}

\section{Feferman-Vaught decompositions for $\tsig{n}$ and $\tpi{n}$}\label{section:FO-decomposition}

In this section, we look at the classes $\tsig{n}$ and $\tpi{n}$ as defined in Section~\ref{section:prelims}. Given that these are indeed the special cases of $\tsinf{\kappa, \lambda}$ and $\tpinf{\kappa, \lambda}$ when $\lambda < \kappa = \omega$, Theorem~\ref{thm:gen-decomposition} yields us {\fv} decompositions for the mentioned classes. It turns out we can say further about the computational aspects of the decompositions as well, as the following theorem shows.

\begin{theorem}\label{thm:FO-decomposition}
Let $\mc{L}$ be one of the logics $\tsig{n}[m]$ or  $\tpi{n}[m]$ for $m, n \in \mathbb{N}$. Let $\tau$ be a vocabulary. Then for every $\mc{L}$ formula $\varphi(\bar{x}_1, \bar{x}_2)$ over $\anntau$, there is an $\mc{L}$ reduction sequence $D(\bar{x}_1, \bar{x}_2)$ over $\tau$  such that:
\begin{enumerate}
    \item $D(\bar{x}_1, \bar{x}_2)$ is a Feferman-Vaught decomposition of $\varphi(\bar{x}_1, \bar{x}_2)$ over the annotated disjoint union operation.\label{thm:FO-decomp:existence}
    \item $D(\bar{x}_1, \bar{x}_2)$ can be computed from $\varphi(\bar{x}_1, \bar{x}_2)$ in time $\tower{n, O((n+1) \cdot |\varphi|^2)}$, and the size of $D(\bar{x}_1, \bar{x}_2)$ is $\tower{n, O((n+1) \cdot |\varphi|)}$.\label{thm:FO-decomp:complexity}
\end{enumerate}
\end{theorem}
\begin{proof}
We show the theorem for $\varphi(\bar{x}_1, \bar{x}_2) \in \tsig{n}[m]$; the case when $\varphi(\bar{x}_1, \bar{x}_2) \in \tpi{n}[m]$ can be handled similarly. Consider the reduction sequence $D(\bar{x}_1, \bar{x}_2)$ given by Theorem~\ref{thm:gen-decomposition} for $\varphi(\bar{x}_1, \bar{x}_2)$. Given that $\tsig{n}[m] = \tsinf{\omega, \lambda}[\mu]$ for $\lambda = n$ and $\mu = m$, we have $\rhohat(\tsig{n}[m]) = \tsig{n}[m]$. Then part (\ref{thm:FO-decomp:existence}) of the theorem holds. We now see part (\ref{thm:FO-decomp:complexity}) by observing the inductive construction of $D(\bar{x}_1, \bar{x}_2)$ in the proof of Theorem~\ref{thm:gen-decomposition} in the case when $\lambda, \mu < \kappa = \omega$ and treating $n$ as in the present theorem as $\lambda$ and $m$ as $\mu$.

\noindent \tbf{Base case}: Let us look at the (outer) base case (case (A)) in the proof of Theorem~\ref{thm:gen-decomposition}. Here $\lambda = 0$ and $\varphi(\bar{x}_1, \bar{x}_2)$ is a quantifier-free FO formula. Consider the construction of $D(\bar{x}_1, \bar{x}_2)$. We make the following observations. 

\begin{enumerate}
    \item In cases (\ref{atomic:non-annot}) and (\ref{atomic:annot}), the size of $D(\bar{x}_1, \bar{x}_2)$ and the time taken to compute it are both at most some suitably large constant $c > 1$ in all cases. 

    \item In case (\ref{andor}), assume as the structural induction hypothesis, that the time taken to compute $D_k(\bar{x}_{1, k}, \bar{x}_{2, k})$ is at most $\tower{0, c \cdot |\varphi_k|^2}$, and that the size of $D_k(\bar{x}_{1, k}, \bar{x}_{2, k})$ is at most $\tower{0, c \cdot |\varphi_k|}$ for $k \in [2]$. Then the time taken to compute $D(\bar{x}_1, \bar{x}_2)$ is 
    \begin{align*}
        \leq & \sum_{k \in [2]} \mbox{Time taken to compute}~ D_k(\bar{x}_{1, k}, \bar{x}_{2, k}) ~~+\\
        & ~~\mbox{Time taken to write}~D(\bar{x}_1, \bar{x}_2)\\
        \leq & \sum_{k \in [2]} \tower{0, c \cdot |\varphi_k|^2} ~~+~~ \sum_{k \in [2]} \mbox{Size of}~D_k(\bar{x}_{1, k}, \bar{x}_{2, k}) ~~+~~ O(1) \\
        \leq & \sum_{k \in [2]} \tower{0, c \cdot |\varphi_k|^2} ~~+~~ \sum_{k \in [2]} \tower{0, c \cdot |\varphi_k|} ~~+~~ O(1) \\
        \leq & ~\tower{0, c \cdot |\varphi|^2} 
    \end{align*}
    The size of $D(\bar{x}_1, \bar{x}_2)$ is  
    \begin{align*}
        = & ~O(1) + \sum_{k \in [2]} \mbox{Size of}~D_k(\bar{x}_{1, k}, \bar{x}_{2, k})\\
        \leq & ~O(1) + \sum_{k \in [2]} \tower{0, c \cdot |\varphi_i|}\\
        \leq & ~\tower{0, c \cdot |\varphi|}
    \end{align*}
    (as $c$ is a sufficiently large constant).
\end{enumerate}

Before we proceed with the induction, we let $\mc{Q}(\lambda, \mc{L})$ denote the following statement. Recall the statement $\mc{P}(\lambda, \mc{L})$ from the proof of Theorem~\ref{thm:gen-decomposition}.\\

\noindent \vspace{3pt}
\begin{tabular}{p{0.15\textwidth}p{0.77\textwidth}}
$\mc{Q}(\lambda, \mc{L}) ~~\equiv$ & For each $\mc{L}$ formula $\varphi(\bar{x}_1, \bar{x}_2)$ over $\anntau$, there is an $\mc{L}$ reduction sequence $D(\bar{x}_1, \bar{x}_2) = (\Delta_1(\bar{x}_1), \Delta_2(\bar{x}_2), \beta)$ over $\tau$ that witnesses $\mc{P}(\lambda, \mc{L})$ for $\varphi(\bar{x}_1, \bar{x}_2)$, and is such that: (i) $D(\bar{x}_1, \bar{x}_2)$ can be computed in time at most $\tower{\lambda, c \cdot (\lambda + 1) \cdot |\varphi|^2}$; (ii) the size of $D(\bar{x}_1, \bar{x}_2)$ at most $\tower{\lambda, c \cdot (\lambda + 1) \cdot |\varphi|}$.
\end{tabular}

\vspace{2pt} \noindent \tbf{Induction}: 
We now look at the induction (case (B)) in the proof of Theorem~\ref{thm:gen-decomposition}; here $\lambda > 0$. In addition to the induction hypothesis assumed in (B), assume for our present proof that for $\mc{L}'$ that is one of the logics $\tsinf{\omega, \lambda'}[\mu']$ or $\tpinf{\omega, \lambda'}[\mu']$, where $\lambda' < \lambda$ and $\mu' < \omega$, the statement $\mc{Q}(\lambda', \mc{L}')$ holds.
We show below that $\mc{Q}(\lambda, \mc{L})$ holds where $\mc{L}$ that is one of the logics $\tsinf{\omega, \lambda}[\mu]$ or $\tpinf{\omega, \lambda}[\mu]$, for an arbitrary $\mu < \omega$. We show this for $\mc{L} = \tsinf{\omega, \lambda}[\mu]$, and by showing the same for $\mc{L} = \tsinf{\omega, \lambda, r}[\mu]$ for all $r$ by a nested induction on $r$ following the corresponding nested induction in the proof of Theorem~\ref{thm:gen-decomposition}. The proof for $\mc{L} = \tpinf{\omega, \lambda}[\mu]$ can be similarly done to complete the present induction.

\vspace{2pt} \noindent \tbf{Nested base case}: This base case is when $r = 0$ whence $\varphi(\bar{x}_1, \bar{x}_2) := \bigwedge_{i \in I}$ $\varphi_i(\bar{x}_{1, i}, \bar{x}_{2, i})$ where the index set $I$ is finite, $\varphi_i$ is a $\tpiinf{\omega, \lambda_i}[\mu]$ formula over $\anntau$ for some $\lambda_i < \lambda (< \omega)$, and $\bar{x}_{j, i}$ is a subtuple of $\bar{x}_j$ for $j \in [2]$ and $i \in I$. For $i \in I$, let $D_i(\bar{x}_{1, i}, \bar{x}_{2, i}) = (\Delta^i_1(\bar{x}_{1, i}), \Delta^i_2(\bar{x}_{2, i}), \beta_i)$ be the $\tpiinf{\omega, \lambda_i}[\mu]$ reduction sequence over $\tau$ witnessing $\mc{Q}(\lambda_i,$ $ \tpiinf{\omega, \lambda_i}[\mu])$ for $\varphi_i(\bar{x}_{1, i}, \bar{x}_{2, i})$, as given by the (outer) induction hypothesis above. We have two cases as in the proof of Theorem~\ref{thm:gen-decomposition}, depending on whether $\lambda = 1$ or $\lambda > 1$. We analyse the latter first, and then the former.

$\mathbf{(a)~\lambda > 1}$: Here $\beta_i$ is of the form 
$\bigwedge_{j \in J_i} (X^{(i, j)}_1 \vee X^{(i. j)}_2)$ where $J_i$ is a finite index set, and  $\Delta^i_k = (\psi^{(i, j)}_k)_{j \in J_i}$ for $k \in [2]$ where $\psi^{(i, j)}_k$ corresponds to $X^{(i, j)}_k$. Recalling the reduction sequence $D(\bar{x}_1, \bar{x}_2)$ for $\varphi(\bar{x}_1, \bar{x}_2)$ as constructed by the proof of Theorem~\ref{thm:gen-decomposition} in equations (\ref{basecase:orofands}), (\ref{basecase:newformulae}) and (\ref{basecase:redseq}), we have the following. Below $J = \{(i, j) \mid i \in I, j \in J_i\},$ the function $f \in \{1, 2\}^J$, and $l \in [2]$.
\begin{align*}
    \beta' := (\bigwedge_{i \in I} \beta_i)  \leftrightarrow \beta''  := \bigvee_{f \in \{1, 2\}^{J}} C_f ~~~~&;~~~~ C_f := \bigwedge_{k \in S_{f, 1}} X^{k}_1 \wedge \bigwedge_{k \in S_{f, 2}} X^{k}_2\\
    \xi_{f, l}(\bar{x}_l) := \bigwedge\limits_{\substack{k \in S_{f, l}\\k = (i, j)}} \psi^k_l(\bar{x}_{l, i})~~~~ &; ~~~~S_{f, l} = \{p \in J \mid f(p) = l\}~~\mbox{for}~l \in [2]\\
    \Delta_l(\bar{x}_l) = (\xi_{f, l}(\bar{x}_l))_{f \in \{1, 2\}^J} ~~~~&;~~~~ \beta := \bigvee_{f \in \{1, 2\}^J} (Y_{f, 1} \wedge Y_{f, 2})
\end{align*}

That $D(\bar{x}_1, \bar{x}_2)$ witnesses $\mc{P}(\lambda, \tsinf{\omega, \lambda}, 0[\mu])$ for $\varphi(\bar{x}_1, \bar{x}_2)$ is already shown in the proof of Theorem~\ref{thm:gen-decomposition}. Towards the size of $D(\bar{x}_1, \bar{x}_2)$, we first observe that for $f \in \{1, 2\}^J$, every pair $(\xi_{f, 1}, \xi_{f, 2})$ corresponds to a unique  subset of the set $\{\psi^k_l \mid k \in J, l \in [2]\}$; the latter set is the same as $\bigcup_{i \in I, l \in [2]} \Delta^i_l$ viewing  $\Delta^i_l$ as a set (instead of as a sequence) of its constituent formulas. Then the size of the pair $(\xi_{f, 1}, \xi_{f, 2})$ is at most the size of $\bigcup_{i \in I, l \in [2]} \Delta^i_l$ which is at most $\sum_{i \in I} |D_i|$ where $|D_i|$ denotes the size of $D_i$. Also since the size of $J_i$ is at most the size of $D_i$, the size of $J$, which is $\sum_{i \in I} |J_i|$, is at most $\sum_{i \in I} |D_i|$. Using these observations and the induction hypothesis, we have the following.
        \begin{equation}
        \begin{aligned}\label{sum-of-D-is}
        \sum_{i \in I} |D_i| & \leq 
        \sum_{i \in I} \tower{\lambda_i, c \cdot (\lambda_i + 1) \cdot |\varphi_i|} \\
        & \leq \sum_{i \in I} \tower{\lambda - 1, c \cdot \lambda \cdot |\varphi_i|} \\ 
        &  \leq \tower{\lambda - 1, c \cdot \lambda \cdot \sum_{i \in I} |\varphi_i|} \\
        & \leq \tower{\lambda - 1, c \cdot \lambda \cdot |\varphi|} \\
        \end{aligned}
        \end{equation}
        \begin{align*}
        |\Delta_1(\bar{x}_1)| + |\Delta_2(\bar{x}_2)|   \leq & \sum_{f \in \{1, 2\}^J} \mbox{Size of}~(\xi_{f, 1}, \xi_{f, 2}) \\
        \leq & ~|\{1, 2\}^J| \cdot \sum_{i \in I} |D_i|\\
        \leq & ~2^{|J|} \cdot \sum_{i \in I} |D_i|\\
        \leq & ~2^{\sum_{i \in I} |D_i|} \cdot \sum_{i \in I} |D_i|\\
        \end{align*}
        For the size of $\beta$, we observe that since there are at most $2 \cdot |\{1, 2\}^J|$ variables $Y_{f, l}$, the number of bits needed to represent any of these variables is at most $\log (2 \cdot |\{1, 2\}^J|)$. Then 
        
        \begin{align*}
            |\beta| \leq & \sum_{f \in \{1, 2\}^J}  \cdot \mbox{Size of}~(Y_{f, 1} \wedge Y_{f, 2})\\
            \leq & \sum_{f \in \{1, 2\}^J} 3 \cdot \log (2 \cdot |\{1, 2\}^J|)\\
            \leq & ~2^{|J|} \cdot 3 \log 2^{|J|+1}\\
            \leq & ~6 \cdot 2^{\sum_{i \in I} |D_i|} \cdot \sum_{i \in I} |D_i|
        \end{align*}
Then the total size of $D(\bar{x}_1, \bar{x}_2)$ is
        \begin{align*}
            \leq &~\mbox{Sum of the sizes of}~\Delta_1(\bar{x}_1), \Delta_2(\bar{x}_2)~\mbox{and}~\beta ~~+~~\mbox{O(1) (for the delimiters)}\\
            \leq & ~7 \cdot 2^{\sum_{i \in I} |D_i|} \cdot \sum_{i \in I} |D_i| ~~+ ~~O(1)\\ 
            \leq & ~8 \cdot 2^{\tower{\lambda - 1, c \cdot \lambda \cdot |\varphi|}} \cdot \tower{\lambda - 1, c \cdot \lambda \cdot |\varphi|} ~~~~\mbox{(from}~(\ref{sum-of-D-is}))\\
            \leq & ~8 \cdot \tower{\lambda, c \cdot \lambda \cdot |\varphi|} \cdot \tower{\lambda - 1, c \cdot \lambda \cdot |\varphi|}\\
            \leq & ~\tower{\lambda, c \cdot (\lambda + 1) \cdot |\varphi|}
        \end{align*}
        For the time taken to compute $D(\bar{x}_1, \bar{x}_2)$, we observe that there is no need to explicitly generate $\beta''$; we can directly write out the pair $(\xi_{f, 1}, \xi_{f, 2})$ by performing $|\{1, 2\}^J|$ many passes over the formulae of $D_i$ for $i \in I$, and extracting out the  relevant $\psi^k_j$'s in each pass. That would give us the sequences $\Delta_l(\bar{x}_l)$ for $l \in [2]$. Finally we directly write out $\beta$ by introducing the new variables $Y_{f, l}$ since we already know $J$ by a single pass over all the $D_i$s. The total time taken to generate $D(\bar{x}_1, \bar{x}_2)$ is thus
        \begin{align*}
            \leq & ~\sum_{i \in I}\mbox{Time taken to compute}~D_i ~~+\\
            & ~\mbox{Time taken for}~|\{1, 2\}^J|~\mbox{passes over the}~D_i\mbox{s to get}~\Delta_1(\bar{x}_1)~\mbox{and}~\Delta_2(\bar{x}_2)~~+\\
            &~\mbox{Time taken to write~}\beta \\
        \end{align*}
        \begin{align*}
            \leq &~ \sum_{i \in I} \tower{\lambda_i, c \cdot (\lambda_i+1) \cdot |\varphi_i|^2} ~~+\\
            & ~~d \cdot |\{1, 2\}^J|\cdot \sum_{i \in I} |D_i| ~~+~~ d \cdot 6  \cdot 2^{\sum_{i \in I} |D_i|} \cdot \sum_{i \in I} |D_i|~~~\mbox{(for some constant}~d > 0)\\
            \leq &~ \sum_{i \in I} \tower{\lambda - 1, c \cdot \lambda \cdot |\varphi_i|^2}  ~~+~~ 7 \cdot d \cdot 2^{\sum_{i \in I} |D_i|} \cdot \sum_{i \in I} |D_i|\\
            \leq &~ \tower{\lambda - 1, c \cdot \lambda \cdot \sum_{i \in I} |\varphi_i|^2}  ~~+~~ 7 \cdot d \cdot \tower{\lambda, c \cdot (\lambda + 1) \cdot |\varphi|}\\ 
            \leq &~ \tower{\lambda - 1, c \cdot \lambda \cdot |\varphi|^2}  ~~+~~ 7 \cdot d \cdot \tower{\lambda, c \cdot (\lambda + 1) \cdot |\varphi|}\\ 
            \leq &~ \tower{\lambda, c \cdot (\lambda + 1) \cdot |\varphi|^2}~~~~~~\mbox{(since}~c~\mbox{is sufficiently large)} \\
         \end{align*}

$\mathbf{(b)~ \lambda = 1}$: Following the corresponding case in the proof of Theorem~\ref{thm:gen-decomposition}, we observe that each $\varphi_i$ is a quantifier-free formula over $\anntau$ and hence $\beta_i$ constructed inductively for $\varphi_i$ need not be structured as an AND of ORs as we had in the case when $\lambda > 1$. A pre-processing to bring $\beta_i$ to this form as done in the proof of Theorem~\ref{thm:gen-decomposition}, runs the risk, for our computational result, of introducing an extra exponential in the time taken to compute $D(\bar{x}_1, \bar{x}_2)$ as well as the size of $D(\bar{x}_1, \bar{x}_2)$, since the AND to OR conversion would be followed by an OR to AND conversion of the formula $\beta'$ which is the conjunction of the pre-processed $\beta_i$s. To avoid this additional exponential, we provide an alternate route to handling this case (even in the proof of Theorem~\ref{thm:gen-decomposition} for this case) as we explain below.

We first recall that $\varphi(\bar{x}_1, \bar{x}_2) := \bigwedge_{i \in I}$ $\varphi_i(\bar{x}_{1, i}, \bar{x}_{2, i})$ where the index set $I$ is finite, $\varphi_i$ is a $\tpiinf{\omega, 0}[0]$ formula over $\anntau$, and $\bar{x}_{j, i}$ is a subtuple of $\bar{x}_j$ for $j \in [2]$ and $i \in I$. For $i \in I$, the outer induction hypothesis yields a $\tpinf{\omega, 0}[0]$ reduction sequence $D_i(\bar{x}_{1, i}, \bar{x}_{2, i}) = (\Delta^i_1(\bar{x}_{1, i}), \Delta^i_2(\bar{x}_{2, i}), \beta_i)$  over $\tau$ witnessing $\mc{Q}(\lambda_i,$ $ \tpiinf{\omega, 0}[0])$ for $\varphi_i(\bar{x}_{1, i}, \bar{x}_{2, i})$. Let the set of variables appearing in $\beta_i$ be $\mc{X}_i = \{X^{(i, k)}_l \mid k \in J_i, l \in [2]\}$ for a finite set $J_i$, and let $\Delta^i_l(\bar{x}_{l, i}) = (\psi^{(i, k)}_l(\bar{x}_{l, i}))_{k \in J_i}$ for $i \in I, l \in [2]$.

We construct the formula $\beta'$ as $\beta' := \bigwedge_{i \in I} \beta_i$. Writing this formula as an OR of ANDs, we have that
\begin{equation}\label{basecase-comp:orofands}
    \begin{aligned}
    \beta' \leftrightarrow \beta''  := \bigvee_{p \in [N]} C_p ~~~~\mbox{where}~~~~ 
     C_p := (\bigwedge_{i \in I} \bigwedge_{k \in S^i_{p, 1}} X^{(i, k)}_1) \wedge (\bigwedge_{i \in I} \bigwedge_{k \in S^i_{p, 2}} X^{(i, k)}_2)
    \end{aligned}        
\end{equation}
    
    Above $N$ denotes the number of conjuncts in $\beta''$ (which is in disjunctive normal form),  the sets $S^i_{p, l}$ (which could be overlapping and some empty) are such that $\mc{X}_i = \bigcup_{p \in [N], l \in [2]} S^i_{p, l}$. We now define the formulae $\xi_{p, l}(\bar{x}_l)$ for $p \in [N]$ and $l \in [2]$ as below.
    \begin{align}\label{basecase-comp:newformulae}
        \xi_{p, l}(\bar{x}_l) & := \bigwedge_{i \in I} \bigwedge_{k \in S^i_{p, l}}  \psi^{(i, k)}_l(\bar{x}_{l, i})
                                            \end{align}
    In the event that $\bigcup_{i \in I} S^i_{p, l} = \emptyset$, we put $\xi_{p, l}(\bar{x}_l) := \true$. Let $Y_{p, l}$ be a new propositional variable for $p \in [N]$ and $l \in [2]$. Consider the reduction sequence $D(\bar{x}_1, \bar{x}_2) = (\Delta_1(\bar{x}_1), \Delta_2(\bar{x}_2), \beta)$ where 
    \begin{align}\label{basecase-comp:redseq}
        \Delta_l(\bar{x}_l) & = (\xi_{p, l}(\bar{x}_l))_{p \in [N]} ~~~~ ;~~~~ \beta := \bigvee_{p \in [N]} (Y_{p, 1} \wedge Y_{p, 2})
    \end{align}
    
    We claim that $D(\bar{x}_1, \bar{x}_2)$ witnesses $\mc{Q}(\lambda, \tsinf{\omega, \lambda, 0}[0])$ for $\varphi(\bar{x}_1, \bar{x}_2)$. Firstly, each of the formulae $\xi_{p, l}$ is a conjunction of quantifier-free formulae over $\tau$, and hence belongs to $\tsinf{\omega, \lambda, 0}[0]$ over $\tau$ (since $\lambda = 1$); then $D(\bar{x}_1, \bar{x}_2)$ is a $\tsinf{\omega, \lambda, 0}[0]$ reduction sequence over $\tau$. Next, $\beta$ is indeed without negations and is a finite OR of conjuncts of the form required by $\mc{P}(\lambda, \tsinf{\omega, \lambda, 0}[0])$. Finally, that $D(\bar{x}_1, \bar{x}_2)$ is a Feferman-Vaught decomposition of $\varphi(\bar{x}_1, \bar{x}_2)$ can be shown entirely analogously as in the nested base case in the proof of Theorem~\ref{thm:gen-decomposition}. These facts show that $D(\bar{x}_1, \bar{x}_2)$ witnesses $\mc{P}(\lambda, \tsinf{\omega, \lambda, 0}[0])$ for $\varphi(\bar{x}_1, \bar{x}_2)$. We now show below that the time taken to compute $D(\bar{x}_1, \bar{x}_2)$ and the size of $D(\bar{x}_1, \bar{x}_2)$ are as required by $\mc{Q}(\lambda, \tsinf{\omega, \lambda, 0}[0])$ to complete the (present) nested base case analysis. We do our computations analogously as done above in the $\lambda > 1$ case.
    
    For the size, we observe that every pair $(\xi_{p, 1}, \xi_{p, 2})$ corresponds to a unique subset of the set $\bigcup_{i \in I, l \in [2]} \Delta^i_l$, so that the size of $(\xi_{p, 1}, \xi_{p, 2})$ is at most $\sum_{i \in I} |D_i|$. Also $N$ is at most $2^{|\mc{X}|}$ where $\mc{X} = \bigcup_{i \in I} \mc{X}_i$ and the size of each $\mc{X}_i$ is at most $|D_i|$. Using these observations and the induction hypothesis, and nearly the same calculations as in the $\lambda > 1$ case, we have the following.
    \begin{equation*}
        \begin{split}
            \sum_{l \in [2]} |\Delta_l(\bar{x}_l)| & \leq 2^{\sum_{i \in I} |D_i|} \cdot \sum_{i \in I} |D_i|\\
            |\beta| & \leq 6 \cdot 2^{\sum_{i \in I} |D_i|} \cdot \sum_{i \in I} |D_i|\\
            |D(\bar{x}_1, \bar{x}_2)| & \leq \tower{1, c \cdot 2 \cdot |\varphi|}
        \end{split}
        \quad \quad
        \begin{split}
            \sum_{i \in I} |D_i| & \leq \tower{0, c \cdot |\varphi|}\\
            8 \cdot 2^{\sum_{i \in I} |D_i|} \cdot \sum_{i \in I} |D_i| & \leq \tower{1, c \cdot 2 \cdot |\varphi|}\\
            \phantom{shubha} & \phantom{dinam}
        \end{split}
    \end{equation*}
    
    For the time taken to compute $D(\bar{x}_1, \bar{x}_2)$, we observe that as opposed to the $\lambda > 1$ case, we would need to  generate $\beta''$ to be able to know the number $N$ and the individual conjuncts $C_l$. The time taken to do this is (singly) exponential in the sum of the sizes of the $\beta_i$s, which in turn is at most exponential in the sum of the sizes of the $D_i$s. Once $\beta''$  is obtained, generating each pair $(\xi_{p, 1}, \xi_{p, 2})$ takes a single pass over all the $D_i$s taken together. That would give us the sequences $\Delta_l(\bar{x}_l)$ for $l \in [2]$. Finally we directly write out $\beta$ by introducing the new variables $Y_{p, l}$. Recalling that $N \leq 2^{\sum_{i \in I} |D_i|}$, the total time taken to generate $D(\bar{x}_1, \bar{x}_2)$ is 
        \begin{align*}
            \leq & ~\sum_{i \in I}\mbox{Time taken to compute}~D_i ~~+~~ \mbox{Time taken to obtain}~\beta''~~+\\
            & ~\mbox{Time taken to otain} ~\Delta_1(\bar{x}_1)~\mbox{and}~\Delta_2(\bar{x}_2)~~+~~\mbox{Time taken to write~}\beta \\
            \leq &~ \sum_{i \in I} \tower{0, c \cdot |\varphi_i|^2} ~~+~~2^{\sum_{i \in I} |D_i|} ~~+\\
            &~~ d \cdot N \cdot \sum_{i \in I} |D_i| ~~+~~ d \cdot 6 \cdot 2^{\sum_{i \in I} |D_i|} \cdot \sum_{i \in I} |D_i|~~~\mbox{(for some constant}~d > 0)\\
        \end{align*}
        \begin{align*}    
            \leq &~\tower{0, c \cdot \sum_{i \in I} |\varphi_i|^2}  ~~+~~ 8 \cdot d \cdot 2^{\sum_{i \in I} |D_i|} \cdot \sum_{i \in I} |D_i|~~~~~~~~~~~~~~~~~~~~~~~~~~~~~~~\\
            \leq &~\tower{0, c \cdot \sum_{i \in I} |\varphi_i|^2}  ~~+~~ 8 \cdot d \cdot \tower{1, c \cdot 2 \cdot |\varphi|}\\
            \leq &~ \tower{1, c \cdot 2 \cdot |\varphi|^2}~~~~~\mbox{(since}~c~\mbox{is sufficiently large)}\\
        \end{align*}
\vspace{2pt}\noindent\tbf{Nested induction}: 
Assume as the nested induction hypothesis that $\mc{Q}(\lambda, $ $\tsinf{\omega, \lambda, r}[\mu])$ holds for $r = r_0 \ge 0$. Consider a formula $\varphi(\bar{x}_1, \bar{x}_2)$ of $\tsiginf{\omega, \lambda,  r_0+1}[\mu]$ over $\anntau$, given by $\varphi(\bar{x}_1, \bar{x}_2) := \exists z \varphi_1(\bar{x}_1, \bar{x}_2, z)$ where $\varphi_1$ is a formula of $\tsiginf{\omega, \lambda, r_0}[\mu]$ over $\anntau$. Corresponding to the two views of $\varphi_1(\bar{x}_1, \bar{x}_2, z)$ as $\varphi_1(\bar{x}_1 \cdot z, \bar{x}_2)$ and $\varphi_1(\bar{x}_1, \bar{x}_2 \cdot z)$, we have by the nested induction hypothesis, that there exist $\tsiginf{\omega, \lambda, r_0}[\mu]$ reduction sequences $D_1(\bar{x}_1 \cdot z, \bar{x}_2) = (\Delta^1_1(\bar{x}_1 \cdot z), \Delta^1_2(\bar{x}_2), \beta_1)$ and $D_2(\bar{x}_1, \bar{x}_2 \cdot z) = (\Delta^2_1(\bar{x}_1), \Delta^2_2(\bar{x}_2 \cdot z), \beta_2)$ over $\tau$ witnessing $\mc{Q}(\lambda, \tsinf{\omega, \lambda, r_0})$ resp. for $\varphi(\bar{x}_1 \cdot z, \bar{x}_2)$ and $\varphi(\bar{x}_1, \bar{x}_2 \cdot z)$. As in the proof of Theorem~\ref{thm:gen-decomposition}, let $\Delta^1_1(\bar{x}_1 \cdot z) = (\psi^{(i, 1)}_1(\bar{x}_1 \cdot z))_{i \in I_1}, \Delta^1_2 = (\psi^{(i, 1)}_2(\bar{x}_2))_{i \in I_1}, \Delta^2_1 = (\psi^{(i, 2)}_1(\bar{x}_1))_{i \in I_2},$ and $\Delta^2_2 = (\psi^{(i, 2)}_2(\bar{x}_2 \cdot z))_{i \in I_2}$. Let $\beta_j := \bigvee_{i \in I_j} (X^{(i, j)}_1 \wedge X^{(i, j)}_2)$, where $X^{(i, j)}_l$ corresponds to the formula $\psi^{(i, j)}_l$ for $i \in I_j, l \in [2]$.

We recall below the reduction sequence $D(\bar{x}_1, \bar{x}_2)$ for $\varphi(\bar{x}_1, \bar{x}_2)$ as constructed by the proof of Theorem~\ref{thm:gen-decomposition} in equations (\ref{induction:newformulae}) and (\ref{induction:redseq}).

\begin{equation*}
    \begin{split}
        \xi^{(i, 1)}_1(\bar{x}_1) &  :=  \exists z \psi^{(i, 1)}_1(\bar{x}_1, z)\\
        \xi^{(i, 2)}_1(\bar{x}_1) &  :=   \psi^{(i, 2)}_1(\bar{x}_1)\\
    \end{split}
    \quad  \quad 
    \begin{split}
        \xi^{(i, 1)}_2(\bar{x}_2) &  :=  \psi^{(i, 1)}_2(\bar{x}_2)\\
        \xi^{(i, 2)}_2(\bar{x}_2) &  :=   \exists z \psi^{(i, 2)}_2(\bar{x}_2, z)\\
    \end{split}
\end{equation*}
\begin{align*}
    \Delta_l(\bar{x}_l) & = (\xi^{(i, 1)}_l)_{i \in I_1} \cdot (\xi^{(i, 2)}_l)_{i \in I_2} ~~~~ ~~~~ \beta := \bigvee_{j \in [2]} \bigvee_{i \in I_j} (Y^{(i, j)}_1 \wedge Y^{(i, j)}_2)
\end{align*}
As the proof of Theorem~\ref{thm:gen-decomposition} shows, the reduction sequence $D(\bar{x}_1, \bar{x}_2)$ witnesses $\mc{P}(\lambda,$ $\tsinf{\omega, \lambda, r_0+1}[\mu])$ for $\varphi(\bar{x}_1, \bar{x}_2)$. We now show that the time taken to compute $D(\bar{x}_1, \bar{x}_2)$ and the size of $D(\bar{x}_1, \bar{x}_2)$ are as required by $\mc{Q}(\lambda, \tsinf{\omega, \lambda, r_0+1}[\mu])$.

For the size of $D(\bar{x}_1, \bar{x}_2)$, we first observe that there is a 1-1 correspondence between the formulae of $D$ and the formulae of the reduction sequences $D_1$ and $D_2$ taken together, and that the size of each formula of $D$ (so $\xi^{(i, j)}_l$) is at most twice the size of the corresponding formula in $D_1$ or $D_2$ (which is $\psi^{(i, j)}_l$). Further we see that the size of $\beta$ (which is ``essentially" $\beta_1 \vee \beta_2$) is at most twice the sum of the sizes of $\beta_1$ and $\beta_2$. Then the size of $D(\bar{x}_1, \bar{x}_2)$ is  
    \begin{align*}
        = &~\mbox{Sum of the sizes of}~\Delta_1(\bar{x}_1), \Delta_2(\bar{x}_2)~\mbox{and}~\beta ~~+~~O(1)~\mbox{(for the delimiters)}\\
        \leq &~2 \cdot \sum_{k \in [2]}~\mbox{Size of}~D_k ~~+~~O(1)\\
        \leq &~2 \cdot 2 \cdot \tower{\lambda, c \cdot (\lambda+1) \cdot |\varphi_1|} ~~+~~O(1)\\
        \leq &~\tower{\lambda, c \cdot (\lambda + 1) \cdot |\varphi|}
    \end{align*}
    The time taken to compute $D(\bar{x}_1, \bar{x}_2)$ is
    \begin{align*}
        \leq & ~\mbox{Time taken to compute}~D_1~\mbox{and}~D_2 + \mbox{Time taken to write out}~D(\bar{x}_1, \bar{x}_2)\\  
        \leq & ~2 \cdot \tower{\lambda, c \cdot (\lambda+1) \cdot |\varphi_1|^2} + d \cdot \tower{\lambda, c \cdot (\lambda+1) \cdot |\varphi|}~~~(\mbox{for some}~d > 1)\\
        \leq & ~\tower{\lambda, c \cdot (\lambda+1) \cdot |\varphi|^2}
    \end{align*}
    This completes the nested induction, and hence the outer induction and the proof. 
\end{proof}

\begin{theorem}\label{thm:FO-genop-decomposition}
Let $\mc{L}$ be one of the logics $\tsig{n}[m]$ or $\tpi{n}[m]$ over a vocabulary $\tau$, for $n, m \ge 0$. Let $\divideontimes$ be a quantifier-free sum-like binary operation on $\tau$-structures. Let $\Xi$ be a quantifier-free definition of $\divideontimes$ and $|\Xi|$ denote the sum of lengths of the formulae of $\Xi$. Then for every $\mc{L}$ sentence $\varphi$, there is an $\mc{L}$ reduction sequence $D$ such that the following hold:
\begin{enumerate}
    \item $D$ is a Feferman-Vaught decomposition of $\varphi$ over $\divideontimes$.\label{thm:FO-genop-decomp:existence}
    \item $D$ can be computed from $\varphi$ in time $\tower{n, O((n+1) \cdot (|\varphi|\cdot|\Xi|^2)^2)}$, and the size of $D$ is $\tower{n, O((n+1) \cdot |\varphi|\cdot|\Xi|^2)}$.\label{thm:FO-genop-decomp:complexity}
\end{enumerate}
\end{theorem}
\begin{proof}
The desired reduction sequence $D$ for $\varphi$ is indeed the reduction sequence for the formula $\psi := \Xi(\varphi)$ as given by Theorem~\ref{thm:FO-decomposition}. That $D$ is a Feferman-Vaught decomposition for $\varphi$ can be shown analogously as in the proof of Theorem~\ref{thm:gen-genop-decomposition}. To see that the size of $D$ and the time taken to compute it are as in the statement of the present theorem, it suffices to show that the size of $\psi$ is $O(|\varphi| \cdot |\Xi|^2)$. To see this, we observe the inductive definition of $\Xi(\varphi)$ as given in Section~\ref{section:prelims}. Let $p$ be the maximum arity of any predicate of $\tau$ and $q$ be the maximum size of any formula in $\Xi$. Let $\gamma$ be a subformula of $\varphi$.
\begin{enumerate}
    \item In the base case of $\gamma$ being an atomic formula or its negation, we see that $|\Xi(\gamma)| \leq 2 \cdot (p + 1) \cdot q$. 
    \item If $\gamma = \circledast_{i \in I} \gamma_i$, then $|\Xi(\gamma)| \leq 1 + \sum_{i \in I} |\Xi(\gamma_i)| \leq  2 \cdot (p + 1) \cdot q  + \sum_{i \in I} |\Xi(\gamma_i)|$. 
    \item If $\gamma = Q \bar{x} \gamma_1$ for $\gamma_1 = \circledast_{i \in I} \gamma_i'$ where $(Q, \circledast) \in \{(\exists, \bigwedge), (\forall, \bigvee)\}$, then $|\Xi(\gamma)| \leq A + B + |\Xi(\gamma_1)|$ where $A$ = length of the string ``$Q \bar{x}$" and $B$ = sum of the lengths of the formulae $\xi_U(x_j)$ for $j \in [r]$ where $\bar{x} = (x_1, \ldots, x_r)$ for $r \ge 0$, plus 2 bits for  two conjunction symbols; so $B \leq (2 \cdot q \cdot r + 2) \leq 2 \cdot (p + 1) \cdot q \cdot r$.
\end{enumerate}

We see then that  in going from $\gamma$ to $\Xi(\gamma)$, at most $2 \cdot (p+1)\cdot q$ symbols are added at each node of the parse tree of $\gamma$. Then for $\gamma := \varphi$ and observing that $p, q \leq |\Xi|$, we have $|\Xi(\varphi)| \leq 2 \cdot (p+1) \cdot q \cdot |\varphi| \leq 2 \cdot 2|\Xi| \cdot |\Xi| \cdot |\varphi| \leq 4 \cdot |\varphi| \cdot |\Xi|^2$, completing the proof. 
\end{proof}

The following corollary can now be proved exactly like Corollary~\ref{cor:gen-composition}.
\begin{corollary}\label{cor:FO-composition}
Let $\mc{L}$ be one of the logics $\tsig{n}[m]$ or $\tpi{n}[m]$ over a vocabulary $\tau$, for $m, n \ge 0$. Given (arbitrary) $\tau$-structures $\str{A}_1$ and $\str{A}_2$, and a quantifier-free sum-like binary operation $\divideontimes$ on $\tau$-structures, the $\mc{L}$ theory of $\str{A}_1 \divideontimes \str{A}_2$ is determined by the $\mc{L}$ theories of $\str{A}_1$ and $\str{A}_2$.
\end{corollary}

We conclude this section with a calculation of a bound on the number of non-equivalent formulae in $\tsig{n}$ and $\tpi{n}$ when the rank and the number of free variables of the formulae in these classes are bounded.

\begin{proposition}\label{prop:size-calculation}
Let $\mc{L}$ be one of the logics $\tsig{n}[m]$ or $\tpi{n}[m]$ over a vocabulary $\tau$, for $n, m \ge 0$. Then up to logical equivalence, for $t \ge 0$, the number of  formulae in $\mc{L}$ whose free variables are among a given $t$-tuple $\bar{x}$ of variables, is $\tower{n+2, (|\tau|+1) \cdot (n+1) \cdot (m+t)^p}$ where $p$ is the maximum arity of the predicates of $\tau$.
\end{proposition}

\begin{proof}
We show using simultaneous induction that the number of non-equivalent $\tsig{n}[m]$ formulae, and the number of non-equivalent $\tpi{n}[m]$ formulae, with free variables among a $t$-tuple $\bar{x}$ are both at most $\tower{n+2, (|\tau|+1) \cdot (n+1) \cdot (m+t)^p}$.

For the base case of $n = 0$, we observe that since the (un-negated) atomic formulae are only of the form $R(x_1, \ldots, x_k)$ for a $k$-ary predicate $R \in \tau \cup \{=\}$, the total number of possible (un-negated) atomic formulae one can construct with at most  $t$ free variables is $\leq (|\tau|+1) \cdot t^p$ where $p$ is the maximum arity of any predicate in $\tau$. Then the total number of possible non-equivalent propositional formulae over the mentioned atomic formulae is $\tower{2, (|\tau|+1) \cdot t^p}$. These propositional formulae being exactly the formulae of $\tsig{0}$ and $\tpi{0}$ up to equivalence, the base case is verified.

Assume as induction hypothesis that the result is true with $n = n_0$ and all $m, t$ and $\bar{x}$. We prove the inductive step for $\tsig{n_0 + 1}[m]$ for any given $m \ge 1$, $t \ge 0$ and $\bar{x}$; the proof for $\tpi{n_0+1}[m]$ is similar. Let $N(u, v)$ be the number of non-equivalent $\tpi{n_0}[u]$ formulae whose free variables are among a $v$-tuple $\bar{y}$ of variables for $v \ge 0$. Then the class of finite conjunctions of formulae in $\tpi{n_0}[u]$ has cardinality at most $2^{N(u, v)}$ up to equivalence. By induction hypothesis, $N(u, v) \leq \tower{n_0+2, (|\tau| + 1) \cdot (n_0 + 1) \cdot (u + v)^p}$. Then from the definition of $\tsig{n}$, we obtain that the number of non-equivalent formulae of $\tsig{n_0+1}[m]$ having free variables among the $t$-tuple $\bar{x}$ is at most
\begin{align*}
     &  \sum\limits_{u = 0}^{u = m} 2^{N(m-u, t+u)}\\
     \leq & \sum\limits_{u = 0}^{u = m} \tower{n_0+3, (|\tau|+1) \cdot (n_0 + 1) \cdot (m+t)^p}\\
     \leq & ~(m+1) \cdot \tower{n_0+3, (|\tau| +1 )\cdot (n_0 + 1) \cdot (m+t)^p}\\
     \leq & ~\tower{n_0+3, (|\tau| +1) \cdot (n_0 + 2) \cdot (m+t)^p}
\end{align*}
This completes the induction and the proof. 
\end{proof}

\section{{\ef} game characterization for equivalence in subclasses of $\tsig{n}$ and $\tpi{n}$}\label{section:EF-game}

Define $\tsig{(n, k)}$, resp. $\tpi{(n, k)}$, to be the subclass of $\tsig{n}$, resp. $\tpi{n}$, consisting of formulae $\varphi(\bar{x})$ in which every quantifier block  on every root-to-leaf path in the parse tree of $\varphi(\bar{x})$ has length equal to $k$. In this section, we provide an {\ef} (EF) game characterization for equivalence with respect to $\tsig{(n, k)}$ (equivalently with respect to $\tpi{(n, k)}$ since the negation of any $\tsig{(n, k)}$ formula is equivalent to a $\tpi{(n, k)}$ formula and vice-versa). Our EF game is, as mentioned in introduction, a two-way version of the $(n, k)$-prefix game defined in~\cite{DS21}. We first provide a characterization of the latter game, and utilize that to characterize equivalence with respect to $\tsig{n}[m]$.

To recall the $(n, k)$-prefix game sketched in the introduction, the game is played on a given pair $((\str{A}_1, \bar{a}_1), (\str{A}_2, \bar{a}_2))$ of structures such that $|\bar{a}_1| = |\bar{a}_2|$, and  the number of rounds in the game is $n$. The Spoiler picks up a $k$-tuple from $(\str{A}_1, \bar{a}_1)$ in the odd rounds, and from $(\str{A}_2, \bar{a}_2)$ in the even rounds. The Duplicator responds in any round with a $k$-tuple in the structure that is not chosen by the Spoiler. Let $\bar{b}_{i, j}$ for $i \in [n]$ and $j \in [2]$ be the tuple chosen in the $i^{\text{th}}$ round in the $j^{\text{th}}$ structure in the above play of the game. The Duplicator is said to win the play if the map $(\bar{a}_1 \mapsto \bar{a}_2) \cdot (\bar{b}_{i, 1} \mapsto \bar{b}_{i, 2})_{i \in [n]}$ is a partial isomorphism between $\str{A}_1$ and $\str{A}_2$. The Spoiler wins the play if the Duplicator does not win the play. The Duplicator (resp. Spoiler) is said to have a winning strategy in the game if she (resp. he) wins every play of the game. (So in particular, the Duplicator has a winning strategy in the 0-round game if $\bar{a}_1 \mapsto \bar{a}_2$ is a partial isomorphism between $\str{A}_1$ and $\str{A}_2$.) 

Denote by $(\str{A}_1, \bar{a}_1) \Rrightarrow_{(n, k)} (\str{A}_2, \bar{a}_2)$ that for every $\tsig{(n, k)}$ formula $\varphi(\bar{x})$ with $|\bar{x}| = |\bar{a}_1|$, it holds that $\str{A}_1 \models \varphi(\bar{a}_1) \rightarrow \str{A}_2 \models \varphi(\bar{a}_2)$. Observe that $(\str{A}_1, \bar{a}_1)$ $\Rrightarrow_{(n, k)} (\str{A}_2, \bar{a}_2)$ is equivalent to saying that for every $\tpi{(n, k)}$ formula $\varphi(\bar{x})$, it holds that $\str{A}_2 \models \varphi(\bar{a}_2) \rightarrow \str{A}_1 \models \varphi(\bar{a}_1)$. We now show the following. The proof is along the same lines as the proof of the standard EF theorem for FO~\cite[Chapter 3]{Lib13} and is provided for completeness.

\begin{theorem}\label{thm:EF-thm-prefix-game}
Let $n, k, r \ge 0$ be given. Let $\tau$ be a vocabulary and for $i \in [2]$, let $\str{A}_i$ be a  $\tau$-structure and $\bar{a}_i$ an $r$-tuple from $\str{A}_i$. Then the following are equivalent:
\begin{enumerate}
    \item The Duplicator has a winning strategy in the $(n, k)$-prefix game on $((\str{A}_1, \bar{a}_1),$ $(\str{A}_2, \bar{a}_2))$.\label{thm:EF-thm-prefix-game:strategy}
    \item $(\str{A}_1, \bar{a}_1) \Rrightarrow_{(n, k)} (\str{A}_2, \bar{a}_2)$ holds.\label{thm:EF-thm-prefix-game:transfer}
\end{enumerate} 
\end{theorem}
\begin{proof}


We show both directions of the equivalence in the theorem statement, by induction on $n$ and $r$ for any fixed value of $k$.

\vspace{2pt} \noindent \ul{(\ref{thm:EF-thm-prefix-game:strategy}) $\rightarrow$ (\ref{thm:EF-thm-prefix-game:transfer})}: 
For the base case of $n = 0$, by the premise (\ref{thm:EF-thm-prefix-game:strategy}), we have that $\bar{a}_1 \mapsto \bar{a}_2$  is a partial isomorphism between $\str{A}_1$ and $\str{A}_2$. Then for any quantifier-free formula $\varphi(\bar{x})$ with $|\bar{x}| = r$, it holds that $\str{A}_1 \models \varphi(\bar{a}_1)$ iff $\str{A}_2 \models \varphi(\bar{a}_2)$. Then (\ref{thm:EF-thm-prefix-game:transfer}) is indeed true.

Assume as induction hypothesis that the stated direction of the theorem holds for $n = n_0$ and all $r \ge 0$ for the value of $k$ fixed at the outset and for any two structures $(\str{A}_1, \bar{a}_1)$ and $(\str{A}_2, \bar{a}_2)$ where $|\bar{a}_1| = |\bar{a}_2| = r$. For $i \in [2]$, let $\str{B}_i$ be a $\tau$-structure and $\bar{b}_i$ be an $r$-tuple from $\str{B}_i$. Suppose the Duplicator has a winning strategy in the 
$(n, k)$-prefix game on $((\str{B}_1, \bar{b}_1),  (\str{B}_2, \bar{b}_2))$ where $|\bar{b}_1| = |\bar{b}_2| = r_0$ for some $r_0 \ge 0$. Consider a $\tsig{(n_0+1, k)}$ formula $\varphi(\bar{x}) := \exists \bar{y} \bigwedge_{i \in I} \psi_i(\bar{x}, \bar{y})$ such that $|\bar{x}| = r, |\bar{y}| = k$, the formula $\psi_i(\bar{x}, \bar{y}) \in \tpi{(n_0, k)}$ for all $i \in I$ where $I$ is a finite index set, and $\str{B}_1 \models \varphi(\bar{b}_1)$. Then there exists a $k$-tuple $\bar{b}_1'$ in $\str{B}_1$ such that $\str{B}_1 \models \bigwedge_{i \in I} \psi_i(\bar{b}_1, \bar{b}_1')$. Let $\bar{b}_2'$ be the $k$-tuple in $(\str{B}_2, \bar{b}_2)$ chosen by the Duplicator according to her winning strategy, in response to the choice of $\bar{b}_1'$ by the Spoiler in $(\str{B}_1, \bar{b}_1)$ in the $(n_0+1, k)$-prefix game on $((\str{B}_1, \bar{b}_1), (\str{B}_2, \bar{b}_2))$. Then the Duplicator continues to have a winning strategy in the $(n_0, k)$-prefix game on $((\str{B}_2, \bar{b}_2, \bar{b}_2'), (\str{B}_1, \bar{b}_1, \bar{b}_1'))$. Hence by induction hypothesis, we get $(\str{B}_2, \bar{b}_2, \bar{b}_2') \Rrightarrow_{(n_0, k)} (\str{B}_1, \bar{b}_1, \bar{b}_1')$. Since $\str{B}_1 \models \psi_i(\bar{b}_1, \bar{b}_1')$ for all $i \in I$, and since $\psi_i(\bar{x}, \bar{y}) \in \tpi{(n_0, k)}$, it follows that $\str{B}_2 \models \psi_i(\bar{b}_2, \bar{b}_2')$ for all $i \in I$. Then $\str{B}_2 \models \varphi(\bar{b}_2)$ completing the induction.

\vspace{2pt}\noindent \ul{(\ref{thm:EF-thm-prefix-game:transfer}) $\rightarrow$ (\ref{thm:EF-thm-prefix-game:strategy})}:
For the base case of $n = 0$, we have for any quantifier-free formula formula $\varphi(\bar{x})$ in NNF with $|\bar{x}| = r$, that $\str{A}_1 \models \varphi(\bar{a}_1) \rightarrow \str{A}_2 \models \varphi(\bar{a}_2)$. This is thus true in particular for atomic formulae  $\varphi(\bar{x})$ and their negations. Then $\bar{a}_1 \mapsto \bar{a}_2$ is indeed a partial isomorphism between $\str{A}_1$ and $\str{A}_2$.

Assume as induction hypothesis that the stated direction of the theorem holds for $n = n_0$ and all $r \ge 0$ for the value of $k$ mentioned at the outset, and for any two structures $(\str{A}_1, \bar{a}_1)$ and $(\str{A}_2, \bar{a}_2)$ with $|\bar{a}_1| = |\bar{a}_2| = r$. We show the inductive step contrapositively. Suppose the Spoiler has a winning strategy in the $(n, k)$-prefix game on $((\str{B}_1, \bar{b}_1),  (\str{B}_2, \bar{b}_2))$ for some $r_0 \ge 0$. In the first round, suppose the Spoiler chooses the $k$-tuple $\bar{b}_1'$ from $(\str{B}_1, \bar{b}_1)$ according to his strategy. Let $S(\bar{x}, \bar{y})$ be the set of all $\tpi{(n_0, k)}$ formulae $\psi(\bar{x}, \bar{y})$ for an $r_0$-tuple $\bar{x}$ and a $k$-tuple $\bar{y}$ such that $\str{B}_1 \models \psi(\bar{b}_1, \bar{b}_1')$. (In model-theoretic parlance, $S(\bar{x}, \bar{y})$ is the $\tpi{(n_0, k)}$-type of the $(r_0+k)$-tuple $(\bar{b}_1, \bar{b}_1')$ in $\str{B}_1$.) Observe that $S(\bar{x}, \bar{y})$ is finite up to equivalence, so let $\widehat{S}(\bar{x}, \bar{y}) \subseteq S(\bar{x}, \bar{y})$  be a finite collection of formulae of $S(\bar{x}, \bar{y})$ such that for every formula in $S(\bar{x}, \bar{y})$, there is an equivalent formula in $\widehat{S}(\bar{x}, \bar{y})$. Consider the $\tsig{(n_0+1, k)}$ formula $\varphi(\bar{x}, \bar{y})$ defined as follows:
\[
\varphi(\bar{x}) := \exists \bar{y} \bigwedge \widehat{S}(\bar{x}, \bar{y}) 
\]
We claim that $\str{B}_1 \models \varphi(\bar{b}_1)$ but $\str{B}_2 \not\models \varphi(\bar{b}_2)$. This would contradict the assumption that $(\str{B}_1, \bar{b}_1) \Rrightarrow_{(n_0+1, k)} (\str{B}_2, \bar{b}_2)$. The first of the mentioned claims is immediate; the tuple $\bar{b}_1'$ can be chosen as the witness in $\str{B}_1$, for the existentially quantified $\bar{y}$ in $\varphi(\bar{x})$. For the second claim, towards a contradiction, suppose $\str{B}_2 \models \varphi(\bar{b}_1)$. Then $\str{B}_2 \models \widehat{S}(\bar{b}_2, \bar{b}_2')$ for some $k$-tuple $\bar{b}_2'$ from $\str{B}_2$. Whereby it follows that 
$(\str{B}_2, \bar{b}_2, \bar{b}_2') \Rrightarrow_{(n_0, k)} (\str{B}_1, \bar{b}_1, \bar{b}_1')$ (since $\widehat{S}(\bar{x}, \bar{y})$ is equivalent to $S(\bar{x}, \bar{y})$, and $S(\bar{x}, \bar{y})$ is the $\tpi{(n_0, k)}$-type of $(\bar{b}_1, \bar{b}_1')$ in $\str{B}_1$). Then by the induction hypothesis, the Duplicator has a winning strategy in the $(n_0, k)$-prefix game on $((\str{B}_2, \bar{b}_2, \bar{b}_2'), (\str{B}_1, \bar{b}_1, \bar{b}_1'))$. This is a contradiction since the Spoiler has a winning strategy in the  $(n_0+1, k)$-prefix game on $((\str{B}_1, \bar{b}_1), (\str{B}_2, \bar{b}_2))$ by assumption, and since $\bar{b}_1'$ is chosen according to this strategy, the Spoiler continues to have a winning strategy  in the $(n_0, k)$-prefix game on $((\str{B}_2, \bar{b}_2, \bar{b}_2'), (\str{B}_1, \bar{b}_1, \bar{b}_1'))$. This completes the induction and the proof. 
\end{proof}

Theorem~\ref{thm:EF-thm-prefix-game} gives a characterization of a one-way transfer of the truth of $\tsig{(n, k)}$ formulae across $\tau$-structures. To get a bi-directional transfer and hence an equivalence of two $\tau$-structures w.r.t. $\tsig{(n, k)}$, we extend the $(n, k)$-prefix game to an immediate two-way version of it, that we call the \emph{$(n, k)$-tree-prefix} game. The game is defined as follows. The game arena is a \emph{set} $\{(\str{A}_1, \bar{a}_1), (\str{A}_2, \bar{a}_2)\}$ of structures where $|\bar{a}_1| = |\bar{a}_2|$, and the game is played for $n$ rounds. In the first round, the Spoiler picks a $k$-tuple  from any one of structures. The Duplicator responds with a $k$-tuple  in the structure not chosen by the Spoiler. In the $i^{\text{th}}$ round for $i > 1$, the Spoiler picks a $k$-tuple  from the structure from which a $k$-tuple was chosen by the Duplicator in the $(i-1)^{\text{th}}$ round. The Duplicator as usual responds (in the $i^{\text{th}}$ round) with a $k$-tuple from the structure not chosen by the Spoiler (in the $i^{\text{th}}$ round). The game concludes after $n$ rounds. Let $\bar{b}_{i, j}$ for $i \in [n]$ and $j \in [2]$ be the tuple chosen in the $i^{\text{th}}$ round in the $j^{\text{th}}$ structure in the above play of the game. The Duplicator is said to win the play if the map $(\bar{a}_1 \mapsto \bar{a}_2) \cdot (\bar{b}_{i, 1} \mapsto \bar{b}_{i, 2})_{i \in [n]}$ is a partial isomorphism between $\str{A}_1$ and $\str{A}_2$. As in the $(n, k)$-prefix game, the Spoiler wins the play if the Duplicator does not win the play, and the Duplicator (resp. Spoiler) has a winning strategy in the game if she (resp. he) wins every play of the game. (Again, the Duplicator has a winning strategy in the 0-round game if $\bar{a}_1 \mapsto \bar{a}_2$ is a partial isomorphism between $\str{A}_1$ and $\str{A}_2$.) The following theorem provides a characterization of equivalence w.r.t. $\tsig{(n, k)}$ in terms of the $(n, k)$-tree-prefix game. Let $(\str{A}_1, \bar{a}_1) \equiv_{(n, k)} (\str{A}_2, \bar{a}_2)$ denote that $(\str{A}_1, \bar{a}_1) \Rrightarrow_{(n, k)} (\str{A}_2, \bar{a}_2)$ and $(\str{A}_2, \bar{a}_2) \Rrightarrow_{(n, k)} (\str{A}_1, \bar{a}_1)$. So $(\str{A}_1, \bar{a}_1) \equiv_{(n, k)} (\str{A}_2, \bar{a}_2)$ is true iff the two structures agree on all  $\tsig{(n, k)}$ formulae  $\varphi(\bar{x})$ with $|\bar{x}| = |\bar{a}_1| (= |\bar{a}_2|)$ iff the  structures agree on all  $\tpi{(n, k)}$ formulae  $\varphi(\bar{x})$ with $|\bar{x}| = |\bar{a}_1|$.

\begin{theorem}\label{thm:EF-thm-tree-prefix-game}
Let $n, k, r \ge 0$ be given. Let $\tau$ be a vocabulary and for $i \in [2]$, let $\str{A}_i$ be a  $\tau$-structure and $\bar{a}_i$ an $r$-tuple from $\str{A}_i$. Then the following are equivalent:
\begin{enumerate}
    \item The Duplicator has a winning strategy in the $(n, k)$-tree-prefix game on $\{(\str{A}_1, \bar{a}_1),$ $(\str{A}_2, \bar{a}_2)\}$.\label{thm:EF-thm-tree-prefix-game:strategy}
    \item $(\str{A}_1, \bar{a}_1) \equiv_{(n, k)} (\str{A}_2, \bar{a}_2)$ holds.\label{thm:EF-thm-tree-prefix-game:transfer}
\end{enumerate} 
\end{theorem}
\begin{proof}
Let $\str{B}_i = (\str{A}_i, \bar{a}_i)$ for $i \in [2]$. We claim that the Duplicator has a winning strategy $S$ in the $(n, k)$-tree-prefix game on $\{\str{B}_1, \str{B}_2\}$ iff she has winning strategies $S_1$ and $S_2$ resp. in the $(n, k)$-prefix games on $(\str{B}_1, \str{B}_2)$ and $(\str{B}_2, \str{B}_1)$. We are then done by Theorem~\ref{thm:EF-thm-prefix-game} and the definition of $\equiv_{(n, k)}$. The forward direction of the claimed equivalence is obvious: $S_1$ and $S_2$ are ``essentially'' just $S$, that is, the response of the Duplicator in each of $S_1$ and $S_2$ to the Spoiler's move in any round is that given by $S$. In the reverse direction, $S$ is just the composition of $S_1$ and $S_2$, that is, if in the first round of the $(n, k)$-tree-prefix game on $\{\str{B}_1, \str{B}_2\}$, the Spoiler plays on $\str{B}_1$, then the Duplicator plays the rest of game according to strategy $S_1$, else she plays according to strategy $S_2$.
It is clear that this strategy $S$ is winning for the Duplicator.  
\end{proof}

Using Theorem~\ref{thm:EF-thm-tree-prefix-game}, we obtain the following corollary that is similar to Corollary~\ref{cor:FO-composition}. The two corollaries are however incomparable since $\tsig{n}[m]$ and $\tpi{n}[m]$ are  incomparable with $\tsig{(n', k)}$ and $\tpi{(n', k)}$ for all (non-zero) values of $n, n', k$ and $m$.

\begin{corollary}\label{cor:FO-composition-using-EF-thm}
Let $\mc{L}$ be one of the logics $\tsig{(n, k)}$ or $\tpi{(n, k)}$ over a vocabulary $\tau$, for $n, k \in \mathbb{N}$. Given (arbitrary) $\tau$-structures $\str{A}_1$ and $\str{A}_2$, and a quantifier-free sum-like binary operation $\divideontimes$ on $\tau$-structures, the $\mc{L}$ theory of $\str{A}_1 \divideontimes \str{A}_2$ is determined by the $\mc{L}$ theories of $\str{A}_1$ and $\str{A}_2$. 
\end{corollary}
\begin{proof}
Let $\str{A}_1', \str{A}_2'$ be $\tau$-structures such that $\str{A}_1 \equiv_{\mc{L}} \str{A}_1'$ and $\str{A}_2 \equiv_{\mc{L}} \str{A}_2'$ where $\equiv_{\mc{L}}$ denotes indistinguishability with respect to all $\mc{L}$ sentences. Then $\str{A}_1 \equiv_{(n, k)} \str{A}_1'$ and $\str{A}_2 \equiv_{(n, k)} \str{A}_2'$. We show that the following holds for all $\mc{L}$ sentences  $\varphi$. 
\begin{align}
    \str{A}_1 \anncupdot \str{A}_2 \models \varphi &~~~\leftrightarrow~~~  \str{A}_1' \anncupdot \str{A}_2' \models \varphi\label{eqn:FO-comp:0}
\end{align}
We can then infer the following equivalences. Let $\Xi$ be a quantifier-free definition of $\divideontimes$.
\[\def\arraystretch{1.3}
\begin{array}{lll}
    & \str{A}_1 \divideontimes \str{A}_2 \models  \varphi & \\
    \leftrightarrow & \str{A}_1 \anncupdot \str{A}_2 \models \Xi(\varphi)&~~~~~(\mbox{by}~(\ref{thm:intp}))\\
    \leftrightarrow & \str{A}_1' \anncupdot \str{A}_2' \models \Xi(\varphi) &~~~~~(\mbox{by}~(\ref{eqn:FO-comp:0})~\mbox{ and since}~\Xi(\varphi) \in \mc{L})\\
    \leftrightarrow & \str{A}_1' \divideontimes \str{A}_2' \models \varphi &~~~~~(\mbox{by}~(\ref{thm:intp}))
\end{array}
\]
We therefore just need to show (\ref{eqn:FO-comp:0}) to complete the proof. In other words, we need to show that for $\str{C} = \str{A}_1 \anncupdot \str{A}_2$ and $\str{C}' = \str{A}'_1 \anncupdot \str{A}'_2$, it holds that $\str{C} \equiv_{(n, k)} \str{C}'$.

Since $\str{A}_1 \equiv_{(n, k)} \str{A}_1'$ and $\str{A}_2 \equiv_{(n, k)} \str{A}_2'$ hold by assumption, we have by Theorem~\ref{thm:EF-thm-tree-prefix-game} that the Duplicator has winning strategies $S_1$ and $S_2$ in the $(n, k)$-tree-prefix game on the sets $\{\str{A}_1, \str{A}_1'\}$ and $\{\str{A}_2, \str{A}_2'\}$ resp. Then the strategy for the Duplicator in the $(n, k)$-tree-prefix game on $(\str{C}, \str{C}')$ is a simple composition of the strategies $S_1$ and $S_2$. Specifically, suppose in a given round, say the $i^{\text{th}}$ for $i \in [n]$, the Spoiler picks up a $k$-tuple $\bar{c}$ from say $\str{C}$. Then $\bar{c} = \bar{a}_1 \cdot \bar{a}_2$ where $\bar{a}_j$ is an $l_j$-tuple from $\str{A}_j$ for $0 \leq l_j \leq k, j \in [2]$ and $l_1+ l_2 = k$. Then for $j \in [2]$, consider the $k$-tuple $\bar{d}_j$ that is an expansion of $\bar{a}_j$ obtained by repeating the last element of $\bar{a}_j$ exactly $k - l_j$ many times. (So for e.g. if $\bar{a}_j = (e_1, e_2, e_3)$ and $k = 5$, then $\bar{d}_j = (e_1, e_2, e_3, e_3, e_3)$.) Treating $\bar{d}_j$ as the move of the Spoiler in $\str{A}_j$ in the $i^{\text{th}}$ round of the $(n, k)$-tree-prefix game on $(\str{A}_j, \str{A}_j')$, let $\bar{d}'_j$ be the $k$-tuple chosen by the Duplicator in $\str{A}_j'$ in the response to $\bar{d}_j$ and according to strategy $S_j$. Since $S_j$ is a winning strategy, two elements of $\bar{d}_j$ are equal iff the corresponding elements in $\bar{d}_j'$ are; then let  $\bar{a}_j'$ be the $l_j$-subtuple of $\bar{d}'_j$ obtained by restricting the latter to its first $l_j$ elements. Now consider the $k$-tuple $\bar{c}'$ of $\str{C}'$ given by $\bar{c}' = \bar{a}_1' \cdot \bar{a}_2'$. This tuple is played by the Duplicator in $\str{C}'$ in response to $\bar{c}$ in $\str{C}$ in the $i^{\text{th}}$ round of the $(n, k)$-tree-prefix game on $(\str{C}, \str{C}')$. 

It is easy to verify that the above described strategy of the Duplicator is indeed winning in the $(n, k)$-tree-prefix game on $\{\str{C}, \str{C}'\}$. Then by Theorem~\ref{thm:EF-thm-tree-prefix-game}, we have $\str{C} \equiv_{(n, k)} \str{C}'$, completing the proof. 
\end{proof}


\section{Conclusion and future work}\label{section:conclusion}

In this paper, we have introduced a ``tree-generalization'' of prefix classes of FO formulae. These classes, denoted $\tsig{n}$ and $\tpi{n}$, are such that (the string corresponding to) any root to leaf path in the parse tree of a $\tsig{n}$ formula is of the form $\exists \cdot (\exists^*\bigwedge \forall^* \bigvee)^* w$, and that in the parse tree of a $\tpi{n}$ formula is of the form  $\forall \cdot  (\forall^*\bigvee \exists^* \bigwedge)^*w$ where $w$ contains no quantifiers. We showed Feferman-Vaught decompositions for formulae in these classes over quantifier-free sum-like operations, that preserve the quantifier-alternation structure as well as bounds on the rank of the formulae, and that are computable in time elementary in the sizes of the formulae. These results are obtained from a more general result that shows Feferman-Vaught decompositions over the aforementioned operations, for formulae of the classes $\tsinf{\kappa, \lambda}$ and $\tpinf{\kappa, \lambda}$ that respectively are infinitary extensions of $\tsig{n}$ and $\tpi{n}$, obtained by allowing conjunctions and disjunctions of arity less than $\kappa$. The decompositions again preserve bounds on the rank and the quantifier-alternation structure of the input formulae. To the best of our knowledge, Feferman-Vaught decompositions have not been studied earlier in the literature for infinitary logics. Further for FO, while rank-preserving decompositions for FO formulae are folklore in the literature, such decompositions preserving the quantifier alternation structure as well, appear to be new. Again, there are only a few results known in the literature showing scenarios where decompositions can be obtained in elementary time.  Our addition to this set of results is via exploiting a syntactic structure, namely low quantifier alternations, that is a feature of the FO descriptions of a wide range of interesting properties and problems in computer science. We finally consider subclasses of $\tsig{n}$ and $\tpi{n}$, denoted $\tsig{(n, k)}$ and $\tpi{(n, k)}$, containing formulae in which all quantifier blocks are of size exactly $k$,  and characterize equivalence with respect to these classes using a two-way variant of the $(n, k)$-prefix game defined in~\cite{DS21}. In doing so, we also characterize when the Duplicator has a winning strategy in the $(n, k)$-prefix game thereby resolving an issue in~\cite{DS21} pointed out in~\cite{FLRV21}.



For future work, we would like to take ahead the results of this paper in various directions as mentioned below. 
\begin{enumerate}
    \item For any fixed vocabulary $\tau$, the classes $\tsinf{\kappa, \lambda}$ and $\tpinf{\kappa, \lambda}$ over $\tau$ clearly stabilize for large enough $\lambda$ keeping $\kappa$ constant, and for large enough $\kappa$ keeping $\lambda$ constant. We are interested in knowing these dependencies between $\kappa$ and $\lambda$. For instance,  the mentioned classes stabilize for $ \lambda \ge f_1(\kappa) = \omega_1$ when $\kappa = \omega_1$, and for $\kappa \ge f_2(\lambda) =\omega$ when $\lambda < \omega$. Knowing these functions $f_1$ and $f_2$ can allow us to define the classes $\tsinf{\kappa, \lambda}$ and $\tpinf{\kappa, \lambda}$ more ``compactly" by putting the bounds on $\lambda$ given by $f_1$ for any fixed $\kappa$, and simultaneously putting bounds on the arities of the conjunctions and disjunctions inductively as given by $f_2(\lambda)$ as $\lambda$ varies. This trimmings would also then reflect in the sentences of the decompositions produced by Theorems~\ref{thm:gen-decomposition} and~\ref{thm:gen-genop-decomposition}. Further, finding the functions $f_1$ and $f_2$ is also involved in an investigation of the sizes of $\tsinf{\kappa, \lambda}$ and $\tpinf{\kappa, \lambda}$ up to equivalence. We would also like to know if the  classes $\tsinf{\kappa, \lambda}$ and $\tpinf{\kappa, \lambda}$  for any fixed $\kappa$ constitute a normal form for the infinitary logic $\mc{L}_{\kappa, \omega}$ just as they are when $\kappa \in \{\omega, \omega_1\}$, and also if $\tsinf{\infty, \infty}$ and $\tpinf{\infty, \infty}$ is a normal form for the logic $\mc{L}_{\infty, \omega}$.

    \item The proofs of Theorems~\ref{thm:gen-decomposition} and~\ref{thm:gen-genop-decomposition} show that while the quantifier alternation structure and bounds on the quantifier rank remain preserved in going from a $\tsinf{\kappa, \lambda}$ or $\tpinf{\kappa, \lambda}$ formula to the sentences of its Feferman-Vaught decomposition, there is a blow-up in the arities of conjunctions and disjunctions. However whether this is blow-up is intrinsically unavoidable is not clear at the present. We would like to investigate this question.

    \item We would like to generalize Theorems~\ref{thm:gen-decomposition} and~\ref{thm:gen-genop-decomposition} to arbitrary vocabularies, so those including constants and function symbols, and those that are not necessarily finite. We would also like to generalize these results to operations that are product-like, and (even sum-like operations) that are not necessarily binary and could possibly even have infinite arities. (Indeed the original decomposition results of Feferman and Vaught~\cite{FV59} were for generalized products of infinitely many structures.) We seek to  investigate applications of the mentioned theorems and the suggested generalizations, to model-theoretic questions about infinite structures, just as their finitary counterparts, namely Theorems~\ref{thm:FO-decomposition} and~\ref{thm:FO-genop-decomposition}, join a family of decomposition theorems that have various applications in computer science.

    \item Given the importance of monadic second order logic (MSO) in algorithmic settings, in particular that many important algorithmic problems like 3-colorability have natural MSO descriptions, we would like to investigate extensions of Theorems~\ref{thm:FO-decomposition} and~\ref{thm:FO-genop-decomposition} to  suitably defined MSO analogues of $\tsig{n}$ and $\tpi{n}$. Once again we observe that even with second order quantifiers, the number of quantifier alternations required to express interesting algorithmic problems, is low, and typically again, just one. (For 3-colorability, the number of second order quantifier alternations is 0, and the total number of quantifier alternations (first and second order quantifiers included) is 1.) 

    \item We would like to obtain an EF game characterization for equivalence in $\tsig{n}[m]$. We propose the following \emph{$\tsig{n}[m]$-game} that we believe could provide the  desired characterization. The game arena is a set $\{\str{A}, \str{B}\}$ of structures. In the first round, the Spoiler picks any structure and a $k_1$-tuple from the structure. The Duplicator responds with a $k_1$-tuple in the structure not chosen by the Spoiler. In the $i^{\text{th}}$ round for $i > 1$, the Spoiler chooses a $k_i$-tuple in the structure from which the Duplicator chose a $k_{i-1}$-tuple in the $(i-1)^{\text{th}}$ round. The Duplicator responds with a $k_i$-tuple in the structure not chosen by the Spoiler. The players must ensure that at the end of any round $r$, the relation $\sum_{i \in [r]} k_i \leq m$ is maintained. If it is impossible to play round $r+1$ ensuring this relation -- in other words, if $\sum_{i \in [r]} k_i = m$ -- then the game concludes after $r$ rounds. Else it concludes after $n$ rounds. The winning condition for the Duplicator in any play of the game is the usual one, that the chosen tuples must form a partial isomorphism between $\str{A}$ and $\str{B}$, and usual again is the notion of a winning strategy for the Duplicator, that she wins every play of the game. 
    \item Finally, we are interested in investigating the model checking problem for $\tsig{n}$ and $\tpi{n}$ over graphs of bounded clique-width. It is known from~\cite{FG04} that under believed complexity theoretic assumptions, there is in general no algorithm that can solve the model checking problem for FO sentences $\varphi$ over graphs of bounded clique-width in time $f(|\varphi|) \cdot n^r$ where $n$ is the number of vertices in the graph, $r \ge 0$ and $f$ is an elementary function of $|\varphi|$ (this holds over even all finite trees which have clique-width at most 3). Intuitively, it seems that the unrestricted number of quantifier alternations in the input FO sentence has a role to play in the mentioned result, given the fact that the number of FO sentences modulo equivalence, of a given rank and arbitrary quantifier alternations, is non-elementary in the rank. In this light, Proposition~\ref{prop:size-calculation} motivates the following question which we would like to answer.

\begin{problem}
For any fixed $k, n \ge 0$, does there exist an algorithm that, given a graph $G$ of clique-width at most $k$ and a $\tsig{n}$ or $\tpi{n}$ sentence $\varphi$, decides whether $G$ satisfies $\varphi$ in time $f_k(|\varphi|) \cdot |G|^r$ where $r \ge 0$ and $f_k$ is an elementary function of $|\varphi|$?
\end{problem}
\end{enumerate}

\textbf{Acknowledgements:}
I thank Julia Knight for the suggestion of generalizing to infinitary languages, the Feferman-Vaught decompositions for $\tsig{n}$ and $\tpi{n}$ proved in the conference version of this paper~\cite[Theorem 3.1]{San21}, as well as for helpful discussions pertaining to the results in Section~\ref{section:gen-decomposition}. I also thank the anonymous referees for their useful comments and for pointing to related results in the literature. 

\bibliographystyle{plain}
\bibliography{refs}

\begin{thebibliography}{10}

\bibitem{AK00}
Chris~J Ash and Julia Knight.
\newblock {\em Computable structures and the hyperarithmetical hierarchy}.
\newblock Elsevier, 2000.

\bibitem{DGKS07}
Anuj Dawar, Martin Grohe, Stephan Kreutzer, and Nicole Schweikardt.
\newblock Model theory makes formulas large.
\newblock In {\em International Colloquium on Automata, Languages, and
  Programming}, pages 913--924. Springer, 2007.

\bibitem{DS21}
Anuj Dawar and Abhisekh Sankaran.
\newblock Extension preservation in the finite and prefix classes of first
  order logic.
\newblock In Christel Baier and Jean Goubault{-}Larrecq, editors, {\em 29th
  {EACSL} Annual Conference on Computer Science Logic, {CSL} 2021, January
  25-28, 2021, Ljubljana, Slovenia (Virtual Conference)}, volume 183 of {\em
  LIPIcs}, pages 18:1--18:13. Schloss Dagstuhl - Leibniz-Zentrum f{\"{u}}r
  Informatik, 2021.

\bibitem{EPR2}
Moshe Emmer, Zurab Khasidashvili, Konstantin Korovin, and Andrei Voronkov.
\newblock Encoding industrial hardware verification problems into effectively
  propositional logic.
\newblock In {\em Proceedings of Formal Methods in Computer Aided Design, FMCAD
  2010, Lugano, Switzerland, October 20 - 23, 2010}, pages 137--144, 2010.

\bibitem{data-exch}
Ronald Fagin, Phokion~G. Kolaitis, Ren{\'e}e~J. Miller, and Lucian Popa.
\newblock Data exchange: semantics and query answering.
\newblock {\em Theor. Comput. Sci.}, 336(1):89--124, 2005.

\bibitem{FLRV21}
Ronald Fagin, Jonathan Lenchner, Kenneth~W. Regan, and Nikhil Vyas.
\newblock Multi-structural games and number of quantifiers.
\newblock In {\em 36th Annual {ACM/IEEE} Symposium on Logic in Computer
  Science, {LICS} 2021, Rome, Italy, June 29 - July 2, 2021}, pages 1--13.
  {IEEE}, 2021.

\bibitem{FV59}
S.~Feferman and R.~Vaught.
\newblock The first order properties of products of algebraic systems.
\newblock {\em Fundamenta Mathematicae}, 47(1):57--103, 1959.

\bibitem{FG04}
Markus Frick and Martin Grohe.
\newblock The complexity of first-order and monadic second-order logic
  revisited.
\newblock {\em Annals of pure and applied logic}, 130(1-3):3--31, 2004.

\bibitem{GJL15}
Stefan G\"{o}ller, Jean-Christoph Jung, and Markus Lohrey.
\newblock The complexity of decomposing modal and first-order theories.
\newblock {\em ACM Trans. Comput. Logic}, 16(1), March 2015.

\bibitem{Gro08}
Martin Grohe.
\newblock Logic, graphs, and algorithms.
\newblock {\em Logic and automata}, 2:357--422, 2008.

\bibitem{Gul10}
Sumit Gulwani.
\newblock Dimensions in program synthesis.
\newblock In {\em Proceedings of the 12th International ACM SIGPLAN Symposium
  on Principles and Practice of Declarative Programming}, PPDP '10, pages
  13--24. ACM, 2010.

\bibitem{Haa14}
Christoph Haase.
\newblock {Subclasses of Presburger arithmetic and the weak EXP hierarchy}.
\newblock In {\em Proceedings of the Joint Meeting of the Twenty-Third EACSL
  Annual Conference on Computer Science Logic (CSL) and the Twenty-Ninth Annual
  ACM/IEEE Symposium on Logic in Computer Science (LICS)}, pages 1--10, 2014.

\bibitem{Har16}
Frederik Harwath.
\newblock A note on the size of prenex normal forms.
\newblock {\em Information Processing Letters}, 116(7):443--446, 2016.

\bibitem{HHS15}
Frederik Harwath, Lucas Heimberg, and Nicole Schweikardt.
\newblock {Preservation and decomposition theorems for bounded degree
  structures}.
\newblock {\em {Logical Methods in Computer Science}}, {Volume 11, Issue 4},
  December 2015.

\bibitem{Hod93}
Wilfrid Hodges.
\newblock {\em Model theory}.
\newblock Cambridge University Press, 1993.

\bibitem{Kar59}
Carol~Ruth Karp.
\newblock {\em Languages with expressions of infinite length}.
\newblock PhD thesis, University of Southern California, 1959.

\bibitem{data-integ}
Maurizio Lenzerini.
\newblock Data integration: A theoretical perspective.
\newblock In {\em Proceedings of the 21st ACM SIGMOD-SIGACT-SIGART Symposium on
  Principles of Database Systems}, PODS '02, pages 233--246. ACM, 2002.

\bibitem{Lib13}
Leonid Libkin.
\newblock {\em Elements of finite model theory}.
\newblock Springer Science \& Business Media, 2013.

\bibitem{Mak04}
Johann~A. Makowsky.
\newblock Algorithmic uses of the {Feferman--Vaught} theorem.
\newblock {\em Annals of Pure and Applied Logic}, 126(1-3):159--213, 2004.

\bibitem{EPR1}
Ruzica Piskac, Leonardo~Mendon{\c{c}}a de~Moura, and Nikolaj Bj{\o}rner.
\newblock Deciding effectively propositional logic using {DPLL} and
  substitution sets.
\newblock {\em J. Autom. Reasoning}, 44(4):401--424, 2010.

\bibitem{RL78}
Cattamanchi~R Reddy and Donald~W Loveland.
\newblock Presburger arithmetic with bounded quantifier alternation.
\newblock In {\em Proceedings of the tenth annual ACM Symposium on Theory of
  computing}, pages 320--325, 1978.

\bibitem{San18}
Abhisekh Sankaran.
\newblock A generalization of the {{\L}o{\'s}-Tarski} preservation theorem --
  dissertation summary.
\newblock {\em arXiv:1811.01014}, 2018.

\bibitem{San21}
Abhisekh Sankaran.
\newblock {Feferman-Vaught} decompositions for prefix classes of first order
  logic.
\newblock In {\em ICLA 2021 Proceedings, 9th Indian Conference on Logic and its
  Applications, March 4 - 7, 2021}, pages 111--116.
  \url{https://www.isichennai.res.in/~sujata/icla2021/proceedings.pdf}, 2021.

\bibitem{MS73}
L.~J. Stockmeyer and A.~R. Meyer.
\newblock Word problems requiring exponential time (preliminary report).
\newblock In {\em Proceedings of the Fifth Annual ACM Symposium on Theory of
  Computing}, STOC '73, page 1–9, New York, NY, USA, 1973. Association for
  Computing Machinery.

\bibitem{Tho97}
Wolfgang Thomas.
\newblock Languages, automata, and logic.
\newblock In {\em Handbook of formal languages}, pages 389--455. Springer,
  1997.

\bibitem{Wan94}
Egon Wanke.
\newblock k-nlc graphs and polynomial algorithms.
\newblock {\em Discrete Applied Mathematics}, 54(2-3):251--266, 1994.

\end{thebibliography}
\end{document}